\documentclass[12pt]{article}
\usepackage{latexsym}
\usepackage{amsmath,amssymb}

\usepackage{graphicx}
\vspace{4ex}
\begin{document}

\vspace*{1cm}

\begin{center}
{\Large \bf Multidimensional Baker-Akhiezer Functions and Huygens' Principle}
\\*[4ex] {\bf O.A.Chalykh$^\dagger$, M.V.Feigin$^{\dagger ,\S}$,
A.P.Veselov$^{\ddagger,\star}$} \\*[2ex]
\end{center}

\noindent $^\dagger$
Department of Mathematics and Mechanics,
Moscow State University,\\
Moscow, 119899, Russia

\noindent $^\S$ Independent University of Moscow, Bolshoy Vlasevsky per. 11,
Moscow, 121002, Russia

\noindent $^\ddagger$
Department of Mathematical Sciences,
Loughborough University,\\
Loughborough,  LE11 3TU, UK

\noindent $^\star$ Landau Institute for Theoretical Physics,
Kosygina 2, Moscow, 117940, Russia

\noindent E-mail addresses: chalykh@nw.math.msu.su, mfeigin@dnttm.ru, A.P.Veselov@lboro.ac.uk

\vspace*{1cm} %

\vspace*{1cm}

\noindent {\bf Abstract.} A notion of rational Baker-Akhiezer (BA) function related to a
configuration of hyperplanes in ${\bf C^n}$ is introduced. It is proved
that BA
function exists only for very special configurations (locus configurations),
which satisfy certain overdetermined algebraic system. The 
BA functions satisfy some algebraically integrable Schr\"odinger
equations, so any locus configuration determines such an equation. Some results
towards the classification of all locus configurations are presented.
This theory is applied to the famous Hadamard's problem of description
of all hyperbolic equations satisfying Huygens' Principle. We show that
in a certain class all such equations are related to locus configurations
and the corresponding fundamental solutions can be constructed explicitly
from the BA functions.

\section*{Introduction}

The notion of Baker--Akhiezer function (BA function) has been introduced by
Kri\-che\-ver
\cite{K} in the theory of finite--gap or algebro--geometric solutions of
the nonlinear
PDE's, integrable by the inverse scattering method \cite{DMN}. The BA
function is a
far-going generalisation of the classical function
$$
\psi = \frac{\sigma{(x-z)}}{\sigma{(x)}\sigma{(z)}} e^{\zeta{(z)}x}
$$
well-known as a solution to the classical Lame equation:
$$
L\psi = \lambda\psi,\,\,\, L =  - \frac{d^2}{d x^2} + 2\wp {(x)},\,\,\, \lambda = -
\wp{(z)}.
$$
Here $\sigma, \zeta $ and $\wp$ are classical Weierstrass elliptic functions
(see e.g. \cite{WW}).

In the degenerate case one has the corresponding trigonometric and rational
versions:
$$
\psi_{trig} = (1 - \frac1{k}\cot x) e^{k x},\quad L =  - \frac{d^2}{d x^2}
+ \frac2{\sin^2 {x}},
$$
$$
\psi_{rat} = (1 - \frac1{k x}) e^{k x}, \quad L =  - \frac{d^2}{d x^2} +
\frac2{x^2}.
$$

Certain multidimensional versions of these functions in the rational and
trigonometric cases
have been introduced by Chalykh and Veselov in \cite{ChV1} in the
theory of quantum
Calogero--Moser problem. In this paper we will restrict ourselves by the rational case only.
The construction of \cite{ChV1} (see also \cite{VSCh}) relates such a BA
function $\psi$
to a configuration ${\mathfrak {A}}$ of the hyperplanes $\Pi_\alpha$ in a complex
Euclidean space ${\bf C}^n$
given by the equations $(\alpha, x)= 0$, taken with some multiplicities
$m_{\alpha}\in{\bf Z}_+$.
Here $\alpha \in {\cal A}$, ${\cal A}$ is a finite set of noncollinear vectors.
The function $\psi(k,x)$, $k,x\in{\bf C}^n$ is determined by certain
analytic properties in $k$
(see Section 1) and exists only for very special configurations.

The most important property of the BA function is that it is an eigenfunction
of the multidimensional algebraically integrable Schr\"odinger operator $L$,
which in our case has the form
\begin{equation}
\label{0.3}
L = -\Delta + \sum\limits_{\alpha\in{\cal A}} \frac{m_{\alpha}(m_{\alpha}+1)
(\alpha,\alpha)}{(\alpha,x)^2}
\end{equation}
(see \cite{ChV1,VSCh}).

When ${\mathfrak {A}}$ is a Coxeter configuration, i.e. ${\mathfrak {A}}$ consists of the
reflection hyperplanes
for some finite reflection group $W$ with $W$-invariant multiplicities,
then the corresponding operator
$L$ is the Hamiltonian
of the generalised quantum Calogero--Moser problem (after Olshanetsky
and Perelomov \cite{OP})
with special integer-valued parameters. The existence of the BA function in
this
case was proved in \cite{VSCh} with the help of Heckman's result \cite{H}.

At that time it was believed that the Coxeter case is the only one, when
$\psi$ does exist, but
it turned out to be not the case. The first non-Coxeter examples 
have been found by the authors in \cite{VFCh} (see also
\cite{ChFV2}).

According to the general procedure proposed by Berest and Veselov in \cite{BV}
this led to the new examples of the hyperbolic equations satisfying the
Huygens' Principle
in Hadamard's sense. Motivated by these results Berest and Lutsenko
started
the investigation of the case when the potential depends on two coordinates
only
and found other new examples of the huygensian equations \cite{BL}. Later
Berest proved
\cite{B} that they have actually found all such equations under assumption
that the potential is
homogeneous of degree $(-2)$. Since a generic Berest--Lutsenko potential
could not be described
by the construction \cite{ChV1}, this was the reason for us to revise it.

In the Section 1 we give such a revised definition of the BA function,
which can be derived from the corresponding Schr\"odinger equation and therefore
covers all the possible cases.
It is remarkable that there exists an effective way to check for a given 
configuration whether BA function exists or not. Namely, as we prove in Sections
2 and 3, the following
overdetermined system of algebraic equations is a necessary and sufficient condition for the existence
of the Baker-Akhiezer function: 
\begin{equation}
\label{tog}
\sum_{\beta\in{\cal A} \atop {\beta\ne\alpha}}
\frac{m_\beta(m_\beta +1)(\beta,\beta)(\alpha,\beta)^{2j-1}}{(\beta,x)^{2j+1}}
\equiv 0  \mbox{ on the hyperplane } (\alpha,x)=0
\end{equation}
for each $\alpha\in{\cal A}$ and $j=1,2,\ldots , m_{\alpha}$.

They are equivalent to the vanishing of the first $m_{\alpha}$ odd terms in
the Laurent expansion
of the corresponding potential
\begin{equation}
\label{uu}
u(x) = \sum_{\alpha\in {\cal A}}\frac
{m_\alpha (m_\alpha +1) (\alpha,\alpha)}{(\alpha,x)^{2}}
\end{equation}
at the hyperplane  $(\alpha,x)=0$. Similar characterisation of the rational
finite-gap
potentials in one dimension has been first proposed in the famous paper
\cite{AMM} by Airault,
McKean and Moser, who introduced the term "locus" in this situation.
We will also use this terminology, calling the equations (\ref{tog}) as well as its general 
affine version (see below) as {\it locus equations}.
Duistermaat and Gr\"unbaum \cite {DG} discovered the interpretation of
such equations as a trivial monodromy condition for the corresponding
one-dimensional
Schr\"odinger equation in the complex domain. We give a similar
interpretation for our
locus equations (\ref{tog}) in Section 2.

However, to describe all the configurations, satisfying the locus
equations (\ref{tog})
(the {\it locus configurations}) seems to be a very difficult problem. 
At the moment it is solved only in dimension 2, where the answer is given by the 
Berest--Lutsenko construction. In dimension $n > 2$ all known examples of the locus configurations
are the Coxeter configurations and their special "deformations" \cite{VFCh},\cite{ChFV2}.
In the section 4 we present all the results, which are known in this direction so far.

The generalisation of our construction to the affine configurations of the
hyperplanes
is discussed in the section 5. The potential $u$ and the locus equations in that case have the form:
\begin{equation}
\label{0.4}
u(x)=\sum_{i=1}^K \frac{m_i(m_i+1)(\alpha_i,\alpha_i)}
{((\alpha_i,x) + c_i)^2}
\end{equation}
\begin{equation}
\label{0.5}
\sum_{j\ne i}\frac{
m_j(m_j+1)(\alpha_j,\alpha_j)(\alpha_i,\alpha_j)^{2s-1}}
{((\alpha_j,x) + c_j)^{2s+1}}\equiv 0
\end{equation}
identically on the hyperplane $(\alpha_i,x) + c_i =0$ for all
$i=1,\ldots, K$ and $s=1,\ldots, m_i$.
Unfortunately, so far a little is known
about the affine
locus configurations, which are not linear, i.e. with not all the
hyperplanes to pass through one point.
Apart from the one-dimensional case investigated in \cite{AMM}, \cite{AM}, there are only some
reducible examples discovered by Berest and Winternitz \cite{BW}.
In fact, we show that the classification problem 
for the affine locus configurations can be reduced 
to the linear case (\ref{tog}) by the isotropic projectivisation procedure.

In the last section we discuss the relations of our BA function $\psi$ and
locus configurations
to the Huygens' Principle. The main result says that for any locus configuration
in dimension $n$ 
the corresponding hyperbolic equation
\begin{equation}
\label{hyp}
(\Box_{N+1} + u(x_1,\ldots,x_n))\phi = 0
\end{equation}
satisfies Huygens'
Principle for large enough odd $N$. Conversely, we show that if the equation (\ref{hyp}) 
satisfies Huygens' Principle and all the Hadamard's coefficients are rational functions, then $u(x)$ has a form 
(\ref{0.4}) for some locus configuration.

We conjecture that this construction gives all huygensian equations of the form
$(\Box_{N+1} + u(x_1,\ldots,x_n))\phi = 0$. In the case $n=1$ it is well-known result by Stellmacher and Lagnese \cite{SL}. 
When $n=2$ and $u$ is homogeneous this follows from Berest's theorem \cite{B}. 
The proof of the general case would lead to the solution of the famous Hadamard's problem in the class (\ref{hyp}).

\section* {1. Rational Baker-Akhiezer function related to a configuration of
hyperplanes.}

Let $ { \cal A}$ be a finite set of noncollinear vectors
$\alpha=(\alpha_1,\ldots,\alpha_n)\in { \bf C } ^n$ with multiplicities
$m_ { \alpha}\in { \bf N}$. We will
assume that $ ( \alpha,\alpha ) =\sum_ { i=1}^n \alpha_i^2 \ne 0$.

{\bf Definition.}{ \it A function $\psi ( k, x )
,\,\, k, x\in { \bf C } ^n$ will be called Baker-Akhiezer function (BA
function), if the
following two conditions are fulfilled:

1) $\psi ( k, x)$ has a form
\begin{equation}
\label{1}
\psi ( k, x ) = \frac { P ( k, x ) } { A ( k ) } e^ { ( k, x ) },
\end{equation}
where $A ( k ) = \prod_ { \alpha \in { \cal A }} { ( k,\alpha ) ^ { m_ {
\alpha }}}$, $P ( k, x)$ is a polynomial in $k$ with the highest term $A(k)$;

2) for all $\alpha\in { \cal A}$
\begin{equation}
\label{2}
\partial_\alpha ( \psi ( k, x ) ( k,\alpha ) ^ { m_\alpha } ) =
\partial_\alpha^3 ( \psi ( k, x ) ( k,\alpha ) ^ { m_\alpha } )=\ldots =
\partial_\alpha^ { 2 m_\alpha - 1 } ( \psi ( k, x ) ( k,\alpha ) ^ {
m_\alpha } )\equiv 0
\end{equation}
on the hyperplane $\Pi_\alpha$: $(k, \alpha) = 0$, where $\partial_\alpha= ( \alpha,\frac { \partial} {
\partial k } )$
is the normal derivative for this hyperplane.}

Notice that (\ref{1}) means that $\psi$ is a rational function of $k$ with
the prescribed poles
along the hyperplanes  $\Pi_\alpha, \alpha \in { \cal A }$ and with the
asymptotic behaviour
at infinity:
$$
\psi = \left(1 + o(1)\right) e^{(k,x)}
$$
when $k \to\infty$ along the rays outside the singularities (cf.\cite {K}).

First of all, in the same way as in \cite{ChV1},\cite{VSCh} one can prove
the following

{\bf Theorem 1.1.} {\it If the Baker-Akhiezer function} $\psi$
{\it exists then it is unique and satisfies the algebraically integrable
Schr\"odinger equation}
\begin{equation}
\label{Seq}
 L\psi = -k^{2}\psi,
\end{equation}
{\it where}
\begin{equation}
\label{SCH}
L = -\Delta + \sum\limits_{\alpha\in{\cal A}} \frac{m_{\alpha}(m_{\alpha}+1)
(\alpha,\alpha)}{(\alpha,x)^2}.
\end{equation}

Algebraic integrability of the operator (\ref{SCH}) means that $L$ is a part of a rich (supercomplete) commutative ring of 
partial differential operators (see \cite{VSCh} for precise definitions). 
This ring is described by the following theorem.

{\bf Theorem 1.2.} {\it Let $ { \cal R_A}$ be the ring of polynomials $f ( k)$
satisfying the following properties
\begin{equation}
\label{quasi}
\partial_\alpha f ( k ) = \partial_\alpha^3 f ( k ) =\ldots =
\partial_\alpha^ { 2 m_\alpha - 1 } f ( k ) \equiv 0
\end{equation}
on the hyperplane $ ( \alpha, k ) =0$ for any $\alpha \in { \cal A}$.

If the Baker-Akhiezer function $\psi ( k, x)$ exists then for any polynomial
$f ( k ) \in { \cal R_A}$ there exists some differential operator
$L_f ( x,\frac { \partial} { \partial x } )$
such that
$$
L_f \psi ( k, x ) = f ( k)\psi ( k, x ).
$$
All such operators form a commutative ring isomorphic to
the ring $ {\cal R_A}$.  The Schr\"odinger operator (\ref{SCH})
corresponds to $f(k)=-k^2$.}

We give the proof of these statements in a more general affine situation
in the Section 5.

We should note that there exists the following explicit formula for $L_f$ (due to Yu.Be\-rest \cite{B2}).

{\bf Theorem 1.3.} {\it The commuting partial differential operators $L_f$ for $f\in {\cal R_A}$ are given by the formula

\begin{equation}
\label{bform}
L_f = c_N (ad_L)^N [\hat f(x)],
\end{equation}
where $c_N=(-1)^N/2^NN!$, $N=deg f$, $\hat f$ is the operator
of multiplication by $f(x)$, and $(ad_L)^N$ means the $N$-th iteration of
the standard $ad$-procedure, $ad_A B = AB-BA$}.

The proof follows from the results of the next section (see Corollary 2.5).

We should note that originally in \cite{ChV1} another axiomatics for the $\psi$ - function was proposed.
There was considered a function $\phi(k,x)$ of the form
\begin{equation}
\label{0.1}
\phi ( k, x ) =  { P ( k, x ) } e^ { ( k, x ) },
\end{equation}
where $P(k,x)$, as in (\ref{1}), is a polynomial in $k$ with the highest
term $A(k)$, with the property
\begin{equation}
\label{0.2}
\partial_\alpha ( \phi ( k, x ) ) =
\partial_\alpha^3 ( \phi ( k, x )  )=\ldots =
\partial_\alpha^ { 2 m_\alpha - 1 } ( \phi ( k, x ) ) \equiv 0
\end{equation}
at the hyperplane $\Pi_\alpha$.

Comparing (\ref{0.1}), (\ref{0.2}) with (\ref{1}), (\ref{2}) we see that the difference between these two axiomatics is due to the additional factor
$\prod_ { \beta \ne  \alpha} { ( k,\beta ) ^ 
{ m_ {\beta }}}$ . In the Coxeter situation considered in \cite{ChV1} (see Section 4
below) this factor
is not essential because of its symmetry.

It turns out that
this minor change
makes the axiomatics less restrictive and leads to a richer class of the
integrable
Schr\"odinger operators. We will prove (see Corollary 2.7) 
that if there
exists $\phi$
satisfying the conditions
(\ref{0.1}), (\ref{0.2}) then there exists also the BA function $\psi$ with
the properties (\ref{1}), (\ref{2}) and
in that case
$\psi =\frac {\phi}{A(k)}$. Converse is not true: there are
configurations,
for which $\psi$ does exist but $\phi$ does not (see remark after the proof of 
the Theorem 4.4).

\section* {2. Monodromy and BA functions.}

Let $L = -\Delta + u(x)$ be a Schr\"odinger operator with a meromorphic potential
$u(x)$ having a pole along the hyperplane $\Pi_\alpha: (\alpha,x)=0$,
which is assumed to be non-isotropic: $(\alpha, \alpha)\ne 0$.

We are looking for a formal solution $\phi$ of the Schr\"odinger equation
$
L\phi = \lambda \phi
$
in the form
\begin{equation}
\label{psi2}
\phi (x) =\sum_{s\ge 0}{\phi_s^{(\alpha)}} (\alpha, x)^{\mu + s},
\end{equation}
for some $\mu$, where the coefficients $\phi_s^{(\alpha)}=\phi_s^{(\alpha)}(x^\bot)$ are some analytic functions on the hyperplane
$\Pi_\alpha$,  $x^\bot$ is
orthogonal projection of $x$
onto $\Pi_\alpha$, $\phi_0^{(\alpha)}\neq 0$.

Let's suppose that the equation
$
L\phi = \lambda \phi
$
has a solution of the form (\ref{psi2}) with some $\mu <0$. Then the substitution into the equation gives immediately that the potential $u(x)$ must have a second order pole along $\Pi_\alpha$: the Laurent expansion in the normal direction $\alpha$ has the form 
\begin{equation}
\label{u}
u(x)=  \sum_{k\ge -2} c_k^{(\alpha)} ( \alpha, x ) ^k
\end{equation}
with $c_{-2}^{(\alpha)}=\mu (\mu - 1)(\alpha, \alpha)$.

Moreover, we obtain the following recurrent relations for the coefficients 
$\phi _s^{(\alpha)}$:
\begin{equation}
\label{rec}
(\alpha, \alpha)(\mu(\mu-1)-(\mu +s)(\mu +s-1))\phi_{s} =
(\widetilde\Delta + \lambda)\phi_{s-2} - \sum\limits_{i=-1}^{s-2} c_i \phi_{s-i-2},
\end{equation}
($s= 1, 2, \ldots$), where $\widetilde \Delta$ is the Laplacian $\Delta$ restricted to the hyperplane $\Pi_\alpha$ and we omitted all the indices $\alpha$ in the coefficients.

If $2\mu \notin {\bf Z}$ we can determine all $\phi_s$ from (\ref{rec}) and obtain the solution (\ref{psi2}) starting from an {\it arbitrary} function $\phi_0$ (the same procedure gives also another solution with $\mu'=-1-\mu$).

In the one-dimensional case this is a classical way (going back to Frobenius, see e.g. \cite{Ince}) to construct the basis of 
solutions of the corresponding equation 
\begin{equation}
\label{ode}
-\varphi''+u(x)\varphi = \lambda \varphi
\end{equation}
 in a vicinity of its regular singular point. 
In the case when the equation (\ref{ode}) has no monodromy in the complex domain, i.e. all the 
solutions are single-valued, we have that 

1) $\mu$ must be an integer: $\mu = -m,\,\,m\in {\bf Z}_+$,

2) the first $2m+1$ equations from (\ref{rec}) must be compatible.

\noindent In case if this is true for {\it each} energy level $\lambda$ we will say that the Schr\"odinger operator
 has trivial monodromy. 

In the multidimensional case there exists a generalisation of Frobenius's theory for the partial differential equations with the regular 
singularities in the complex domain (see \cite{O}). For the Schr\"odinger equation with a singularity along a hypersurface the regularity 
condition means that the potential has a second order pole at most.

The considerations above motivate the following 

{\bf Definition}. {\it We say that a Schr\"odinger operator $L = -\Delta + u(x)$ with  meromorphic potential
$u(x)$ having a second order pole along the hyperplane $\Pi_\alpha: (\alpha,x)=0$
has {\it local trivial monodromy} around this hyperplane if

1) the Laurent coefficient $c_{-2}^{(\alpha)}$ in the expansion (\ref{u}) has the form
 $c_{-2}^{(\alpha)}=m_\alpha(m_\alpha +1)(\alpha,\alpha)$
for some $m_\alpha\in{\bf Z}_+$,

2) the system (\ref{rec}) with $\mu=-m_\alpha$ is compatible for any function $\phi_0$ and for all $\lambda \in {\bf C}$.}

{\bf Theorem 2.1.} {\it $L$ has local trivial monodromy around $\Pi_\alpha$
if and only if the
coefficients of the normal Laurent expansion of the potential $u(x)$ near
$\Pi_\alpha$
$$
u(x)=  \sum_{s\ge -2} c_s^{(\alpha)} ( \alpha, x ) ^s
$$
satisfy the following conditions: 
$c_{-2}=m_\alpha(m_\alpha+1)(\alpha,\alpha)$ for some 
$m_\alpha \in {\bf Z}_+$,\ and
\begin{equation}
\label{***}
\mbox{  } c^{(\alpha)}_{-1} = c^{(\alpha)}_1 = c^{(\alpha)}_3 = \ldots =
c^{(\alpha)}_{2 m_\alpha -1}\equiv 0
\mbox{  on   }\Pi_\alpha.
\end{equation}

In that case the Laurent expansions of the corresponding eigenfunctions
$\phi$ (\ref{psi2})
satisfy the conditions
\begin{equation}
\label{ax}
\phi^{(\alpha)}_{1} = \phi^{(\alpha)}_3 = \ldots = \phi^{(\alpha)}_{2
m_\alpha -1}\equiv 0
\mbox{  on  }\Pi_\alpha.
\end{equation}
}

{\bf Proof} is similar to the one-dimensional case considered by
J.Duister\-maat and A.Gr\"un\-baum
\cite{DG}. Let's demonstrate the idea in the simplest case when
$m_\alpha=1$. After substitution (\ref{psi2})
into the Schr\"odinger equation, we deduce that $\mu = 2$ and derive the
following recurrent relations for
$\phi^{(\alpha)}_{k}$:

\begin{equation}
\label{m1}
\left\{
\begin{array}{l}
(-2+c_{-2})\phi_0 = 0 \\
2\phi_1 + c_{-1}\phi_0 = 0\\
2\phi_2+(-\widetilde\Delta -\lambda)\phi_0+c_0\phi_0+c_{-1}\phi_1=0\\
0\phi_3+(-\widetilde\Delta
-\lambda)\phi_1+c_1\phi_0+c_{0}\phi_1+c_{-1}\phi_2=0\\
\ldots
\end{array}
\right.
\end{equation}
where $\widetilde\Delta $ is the Laplacian $\Delta$ restricted to the
hyperplane
$\Pi$ (we omitted all the subindices $\alpha$ in these formulas and
assumed that $(\alpha,\alpha) = 1$). These relations allow one to find all
the coefficients uniquely except $\phi_0$ (which is an arbitrary function)
and $\phi_3$, provided the consistency of the first four equations. From
the first equation it follows that $c_{-2}=2$.  Expressing $\phi_1$ and
 $\phi_2$ from the second and the third equations and substituting them
into the fourth one we arrive at the relation
$$
(-\widetilde\Delta-\lambda)(-\frac12 c_{-1} \phi_0) - \frac12 c_{-1}
(-\widetilde\Delta-\lambda)\phi_0+
(c_1 - c_0 c_{-1} +\frac14 c_{-1}^3)\phi_0 = 0,
$$
which should be valid for all $\phi_{0}$ and $\lambda$. Vanishing of the
leading term in
$\lambda$ gives $c_{-1}\phi_0\equiv 0$, i.e. $c_{-1}\equiv 0$. The relation
reduces after that to $c_1\phi_0 = 0$, thus $c_1\equiv 0$. Notice that as it
follows from the second equation
$\phi_1 = -\frac12 c_{-1}\phi_2\equiv 0$. This completes the proof in the
case when $m_\alpha=1$. In the general
case one should use induction arguments (see \cite{DG}, p.196).

\noindent {\it Remark.} One can consider a more general case, when $u(x)$ has a
singularity along an arbitrary hypersurface
$\varphi(x)=0$. However, analysis of the corresponding relations (\ref{m1})
shows that the hypersurface has to be
a hyperplane (cf. \cite{BV98}).

Now let's consider a Schr\"odinger operator (\ref{0.3}) $L$,
corresponding to some Baker -- Akhiezer function
$\psi$. We claim that such an operator
has local trivial monodromy
around all the singular hyperplanes.
To prove one can consider for a given  $\lambda$  the $(n-1)$ - dimensional
family of the solutions of the Schr\"odinger equation
$$
(L-\lambda)\varphi = 0
$$
of the form $\varphi = \psi(k,x)$ with $k^2 = -\lambda$. They have proper
pole behaviour near the hyperplane
$(\alpha,x)=0$. Unfortunately, $\psi^{(\alpha)}_0$ depends on $k$ and is
not an arbitrary function
on the hyperplane, so we have to present additional arguments. We'll prove a
slightly more general result, which we
will use also in the section 6.

{\bf Theorem 2.2.}
{\it Let the Schr\"odinger operator $L=-\Delta + u(x)$ have an
eigenfunction $\psi(k,x)$
$$
L\psi=-k^2\psi
$$
of the form
$\psi=P(k,x) e^{(k,x)}$, where $P$ is a finite sum of some functions which are
homogeneous in $k$ and meromorphic in $x$. Then the singularities of $u(x)$
are second order poles
located on a union of non-isotropic hyperplanes and $L$ has local trivial monodromy
around these hyperplanes.}

{\bf Proof.} The fact that singularities of $u(x)$ must be located on the
hyperplanes was proved
by Yu.Yu. Berest and A.P. Veselov in
\cite{BV198} under assumption that $P$ is a polynomial in $k$, but  their
proof works also in the case
when $P$ is a finite sum of the homogeneous in $k$ functions. The fact that these 
hyperplanes must be non-isotropic follows from the zero-residue lemma of the
same 
paper \cite{BV198} (see also \cite{BV98}).

Let's now
prove that the conditions
(\ref{***}) are to be satisfied. After a proper choice of orthonormal
basis we may assume
that the hyperplane under consideration has the equation
$x_1=0$, and let's consider the Laurent expansion
for the function $\psi(k,x)$: 
\begin{equation} 
\label{lpsi} 
\psi ( k, x ) = x_1 ^ { -m}\sum\limits_ { i=0}^ {+\infty } \psi_i
 (k,x_2,\ldots,x_n)  x_1^i.  
\end{equation} 
Let's prove first that $m$ has
to be positive. Let $P^0$ be the highest homogeneous term of $P$, then
from the Schr\"odinger equation we have $\sum k_i \partial/\partial x_i
P^0 = 0$. So $P^0(k,x+k t)$ is constant while $t$ varies, hence if $P^0$
vanishes on the hyperplane $x_1=0$, then it vanishes identically.  Thus,
$P^0$ and therefore $\psi$ can not be zero at the hyperplane, so $m$ in
(\ref{lpsi}) must be positive.

Substitution (\ref{lpsi}) to Schr\"odinger equation immediately gives that
$c_{-2}=m(m+1)$ and leads to
the following
recurrence relations:
\begin{equation}
\label{rec1}
(m(m+1)-(j+2-m)(j+1-m))\psi_{j+2} =
(\widetilde\Delta - k^2)\psi_j - \sum\limits_{i=-1}^j c_i \psi_{j-i},
\end{equation}
($j=-1, 0, 1, 2, \ldots$), $\widetilde \Delta = \frac{\partial^2}{\partial
{x_2}^2} + \ldots
+ \frac{\partial^2}{\partial {x_n}^2}$ .
To prove (\ref{***}) let's suppose that
$c_{-1}=c_1=\ldots=c_{2p-3}=0$, but $c_{2p-1}\ne 0$ for some $p<m+1$.
Considering $j=-1, 1, 3, \ldots, 2p-3$ it is easy to see that
$\psi_{1}=\psi_3=\ldots=\psi_{2p-1}=0$. From the form of the function $\psi$ it follows
that $\psi_j=P_j(k,x_2,\ldots,x_n)e^{(\tilde k,\tilde x)}$, where $P_j$ are
finite sum of homogeneous functions in $k$, $\tilde k = (k_2,\ldots, k_n)$,
$\tilde x = (x_2,\ldots, x_n)$. Let $P_j^0$  be the highest homogeneous term of
$P_j$. By induction one can prove that $P_{2j}^0=(-1)^j k_1^{2j}P_{0}^0a_j$ and
$P_{2j-1}^0=(-1)^{j-p} k_1^{2(j-p-1)}P_{0}^0 c_{2p-1}b_j$,
where the constant $a_j>0$ and $b_1=b_2=\ldots=b_p=0$ (by assumption) and
$b_j>0$ for $m\ge j\ge p+1$. Indeed, for $P_{2j}^0$ it follows easily from
the relations
(\ref{rec1}). For $P_{2j-1}^0$ one can use induction arguments similar to
\cite{DG} ( prop. 3.3, p. 196).

Now let's consider the equation (\ref{rec1}) with the resonance value 
$j=2m-1$:
$$
0 =
(\widetilde\Delta - k^2)\psi_{2m-1} - \sum\limits_{i=-1}^{2m-1} c_i
\psi_{2m-1-i}.  $$ Since this holds identically for all $k$ the highest
homogeneous term should vanish.  Simple calculation shows that this term is
equal to 
$$
 -(P_{2m-1}^0 k_1^2+ P_{2m-2p}^0 c_{2p-1}) 
=
(-1)^{m-p+1}k_1^{2(m-p)}P_0^0 (b_m+a_m) c_{2p-1} 
.
$$
Since $b_m+a_m>0$ and $P_0^0\ne 0$ it vanishes only if $c_{2p-1}=0$. This
completes the proof.

It is remarkable that the BA function  turns out to be symmetric with
respect to
$k$ and $x$. For Coxeter configurations this property has been established in
\cite{VSCh}.

{\bf Theorem 2.3.} {\it Baker-Akhiezer function} $\psi(k,x)$ {\it is symmetric
with respect to} $x$ {\it and} $k$:
$\psi(k,x) = \psi(x,k)$.

{\bf Proof.} The idea is to show that $\psi(x,k)$ is also the BA function
and then to use the uniqueness (theorem 1.1).
Let's prove that $\frac{A(x)P(k,x)}{A(k)}$ is a polynomial in $x$ with the
highest
term $A(x)$, where $A(x)$ and $P(k,x)$ are the same as in (\ref{1}).
For that let us consider the conditions (\ref{2}) for $\psi(k,x)$.
They give a linear system for the coefficients
of the polynomial $P$ with the coefficients, which are proportional to the
degrees $ ( \alpha, x)$, $\alpha \in { \cal A}$. Since this system has a
unique
solution,
these coefficients are rational in $x$. Let' denote by  $P_j(k,x)$ the
homogeneous term of $\frac{P(k,x)}{A(k)}$ of degree $-j$ in $k$.
In terms of $P_j(k,x)$ one can rewrite the equation
(\ref{Seq})  in the following recurrent way
$$
L P_j(k,x) = 2\sum_{i=1}^n
k_i \frac{\partial}{\partial {x_i}} P_{j+1},\,\,\,\, P_0 (k,x) = 1.
$$
From this it follows by induction that all the singularities of $\psi(k,x)$ in
$x$ belong to our configuration of the hyperplanes $(\alpha,x)=0$.
Analyzing Laurent expansions for $u(x)$
and $\psi(k,x)$
on these hyperplanes we conclude that $\psi(k,x)$ has a pole of
order $m_\alpha$ along the hyperplanes $ ( \alpha, x ) =0$. All that means
that $A(x)P(k,x)$ is a polynomial in $x$. But from the uniqueness of
BA-function it follows easily that $P_j(k,x)$
is also homogeneous in $x$ with the same degree $-j$. Hence the highest
term in $x$ of the polynomial $A(x)P(k,x)$ is equal to $A(x)A(k)$.
Thus $\psi(x,k)=\frac{A(x)+\ldots}{A(x)}e^{(k,x)}$. Properties of the
Laurent expansions
in $x$ follow immediately from the theorems 2.1, 2.2. So we
have all the conditions for $\psi(x,k)$ to be a BA function. The theorem
is proved.

{\bf Corollary 2.4.} {\it Baker-Akhiezer function} $\psi$ {\it satisfies the
following
 bispectral problem}
\begin{equation}
\label{BP}
L ( x,\frac { \partial} { \partial x }) \psi(k,x) = -k^2 \psi(k,x),\quad
L ( k,\frac { \partial} { \partial k }) \psi(k,x) = -x^2 \psi(k,x),
\end{equation}
{\it where $L$ is the Schr\"odinger operator (\ref{SCH}).}

Now we are able to prove Theorem 1.3.

{\bf Corollary 2.5.} {\it The Baker-Akhiezer function $\psi$ is an eigenfunction of the operator (\ref{bform}) for any $f\in {\cal R_A}$.}

{\bf Proof}. Due to the Theorem 1.2 and to the symmetry of $\psi$ for any  
$f\in {\cal R_A}$ there exists a differential operator 
$A ( k,\frac { \partial} { \partial k })$ such that 
$A ( k,\frac { \partial} { \partial k })\psi = f(x) \psi$. On the other hand,   $L ( x,\frac { \partial} { \partial x }) \psi = -k^2 \psi$. Now we can use the identity (1.8) from \cite{DG} which states in that case that
$$
(ad L)^r(\hat f)[\psi]=(-ad \hat k^2)^r(A)[\psi]
$$
for all $r\in {\bf Z}_+$. For $r=N={\rm ord}A={\rm deg}f$ the differential operator $(-ad \hat k^2)^r(A)$ in the right-hand side has zero order and is, in fact, the operator of multiplication by $cf(k)$ with $c=(-2)^NN!$. This means that 
$\psi$ is an eigenfunction of the operator $(ad L)^r(\hat f)$ with the eigenvalue $cf(k)$. This proves the theorem 1.3.

Now let's explain why the existence of $\phi$ with the properties
(\ref{0.1}), (\ref{0.2})
(our old axiomatics, see section 1) implies the existence of BA
function $\psi$.
This follows from the following general statement, showing that the new
axiomatics is
in some sense the most general one.

Let ${\cal A}$ be any configuration of hyperplanes, $L=-\Delta+u(x)$ be a
corresponding
Schr\"odinger operator, $A(k)=\prod_{\alpha\in{\cal A}}
(\alpha,k)^{m_\alpha}$. Consider
the functions $\varphi$ of the form
\begin{equation}
\label{sol}
\varphi(k,x) = \frac{P(k,x)}{A(k)A(x)}e^{(k,x)},
\end{equation}
$P$ is some polynomial in $k$ and $x$: $P=A(k)A(x)+\ldots$, where dots mean
the terms of lower
order both in $k$ and in $x$.

{\bf Theorem 2.6.} {\it If the Schr\"odinger equation
$L\varphi=-k^2\varphi$ has
a solution $\varphi$ of the form (\ref{sol}) then $\varphi(k,x)$ has to be
BA function.}

{\bf Proof} now is almost evident. Theorems 2.1 and 2.2 provide the conditions
(\ref{2}) for
$\varphi$ in $x$-variable, and it has the required form (\ref{1}) in $x$.
Hence,
$\varphi(x,k)$ is BA function and according to the theorem 2.3
$\varphi(x,k)=\varphi(k,x)$.

{\bf Corollary 2.7.} {\it If a function $\phi$ satisfies the conditions (\ref{0.1})-(\ref{0.2}) then $\psi=A^{-1}(k)\phi$ is the Baker-Akhiezer
function (\ref{1})-(\ref{2}).}

{\bf Proof}. As it follows from the results of the papers \cite{ChV1}, \cite{VSCh}, the function $\phi$ 
must be an eigenfunction of the same equation (\ref{Seq}). Then the arguments we used in the proof of 
the theorem 2.3 show that $\varphi = A^{-1}(k)\phi$ satisfies the conditions of the theorem 2.6 and 
therefore is the Baker-Akhiezer function.

\section* {3. Locus equations and the existence of BA function.}

Let ${\cal A}$, as in Section 1, be a finite set of non-collinear vectors
$\alpha \in { \bf C } ^n$ with given multiplicities $m_ { \alpha}\in {\bf Z}_{+}$, ${\mathfrak {A}}$ be the corresponding configurations of hyperplanes $(\alpha,k)=0$ in ${\bf C}^n$
and $ L = -\Delta + u(x)$ be the Schr\"odinger operator 
with the potential
\begin{equation}
\label{3.0}
u(x) = \sum_{\alpha\in {\cal A}}\frac
{m_\alpha (m_\alpha +1) (\alpha,\alpha)}{(\alpha,x)^{2}}.
\end{equation}

The theorems 2.1 and 2.2 from the previous section imply that if the 
BA function for the configuration ${\mathfrak {A}}$ exists then in the normal 
Laurent expansions (\ref{u}) of the potential $u(x)$ the first odd terms 
$c_{2j-1}^{(\alpha)}$ ($j=1,\ldots , m_{\alpha}$) should vanish identically on the hyperplane 
$(\alpha, x) = 0$. More explicitly, these conditions have the form of the following highly overdetermined algebraic system:
\begin{equation}
\label{togd}
\sum_{\beta\in{\cal A} \atop {\beta\ne\alpha}}\frac
{m_\beta(m_\beta +1)(\beta,\beta)(\alpha,\beta)^{2j-1}}{(\beta,x)^{2j+1}}
\equiv 0  \,\,\mbox{ on the hyperplane } (\alpha,x)=0
\end{equation}
for $j=1,2,\ldots,m_{\alpha}$.
 
We will call the equations (\ref{togd}) as {\it locus equations},
following Airault, McKean and Moser \cite{AMM}, who used this terminology
in one-dimensional case. The configurations ${\mathfrak {A}}$ which satisfy the locus equations we
will call as {\it locus configurations}.

The remarkable fact is that the locus equations (\ref{togd}) are not only necessary, but are also  sufficient for the existence of the BA function. We will give the proof following the paper
\cite{Ch}. 

{\bf Theorem 3.1.} {\it For any locus configuration ${\mathfrak {A}}$ the BA function $\psi(k,x)$  does exist and can be given by the following Berest's
formula:
\begin{equation}
\label{formula}
\psi(k,x)= [(-2)^MM! A(k)]^{-1}
(L+k^2)^M[\prod_{{\alpha}\in {\cal A}} ({\alpha},x)^{m_{\alpha}}exp(k,x)],
\end {equation}
where $M=\sum_{{\alpha}\in {\cal A}} m_\alpha,\,
A(k)=\prod_{{\alpha}\in {\cal A}} ({\alpha},k)^{m_\alpha}$.}

{\bf Proof}. Let's consider the linear space $V$ which consists of
the functions $\phi (x),\, x\in {\bf C^n}$, with the following analytic
properties:
\\
1) $\phi (x)\prod_{\alpha\in{\cal A}} ({\alpha},x)^{m_{\alpha}}$ \,\,
is holomorphic in  ${\bf C^n}$;
\\
2) for each $\alpha\in{\cal A}$ the Laurent expansion (\ref{psi2}) for $\phi$ should
not contain the terms of order  $-m_{\alpha}+2j-1$ ($j=1,\ldots , m_{\alpha})$, i.e.
the conditions (\ref{ax}) hold.

The basic observation is the following

{\bf Lemma.} {\it The space $V$ defined above is invariant under
the Schr\"odinger operator with the potential (\ref{3.0}) provided that the locus conditions (\ref{togd}) are fulfilled.}

It follows easily from the imposed conditions on the Laurent expansions in $\alpha$-direction for $u(x)$ and $\phi \in V$.

Now let's define the functions $\varphi_i \, (i=0,1,\ldots)$ in the following
way: 
$$
\varphi_0 = \prod_{\alpha\in{\cal A}} ({\alpha},x)^{m_{\alpha}}exp(k,x)
$$ 
and
\begin{equation}
\label{phi}
\varphi_{i+1} = (L + k^2) \varphi_i .
\end{equation}
It's obvious that $\varphi_0$ belongs to $V$, hence by the lemma $\varphi_i$
also belongs to $V$. From the definition of these functions and the
property 1 of $V$ it is clear that $\varphi_i$ can be presented in the form
$\varphi_i = R_i(k,x) exp(k,x)$, where
$R_i = Q_i \prod_{\alpha\in{\cal A}} ({\alpha},x)^{-m_{\alpha}}$ for some
polynomial
$Q_i(k,x)$. From (\ref{phi}) it follows that the degrees of the polynomials $Q_i$ in $x$
decrease: ${\rm deg}Q_{i+1}<{\rm deg}Q_i$. Therefore, for some 
$N$ $\varphi_{N}\neq 0$ but $\varphi_{N+1} = (L+k^2)\varphi_N = 0$. Thus, $\phi = \varphi_N$ is an 
eigenfunction for the Schr\"odinger operator $L$. 
Let's prove that $N$ in fact equals to $M=\sum_{{\alpha}\in {\cal A}} m_\alpha$.
If we denote by $R_i^0$ the highest
homogeneous terms of $R_i$ in $x$, we see from (\ref{phi}) that
$$
R_{i+1}^0 = -2\sum_{j=1}^n k_j \partial/\partial x_j\left( R_i^0\right).
$$
From this we obtain immediately that for $i=M=\sum_{\alpha\in {\cal A}}
m_\alpha$
\begin{equation}
\label{R}
R_M^0 = (-2)^MM! \prod_{{\alpha}\in {\cal A}} ({\alpha},k)^{m_{\alpha}}.
\end{equation}
From this we conclude that for $i>M$ $R_i$ (which is polynomial in $k$)
will be of the negative degree in $x$. Thus, it cannot be an eigenfunction for 
the Schr\"odinger operator $L$ because of the following lemma due to F.A.Berezin \cite{Ber}.

{\bf Lemma}. {\it If a quasipolynomial $\psi$ in $k$  $\psi = P(k,x)exp(k,x)$ satisfies the Schr\"odinger equation $(-\Delta + u(x))\psi = -k^2 \psi$ then the highest term in $k$ of
the polynomial $P$ must be polynomial in $x$.}

This contradiction proves that the last non-zero function in the sequence (\ref{phi}) is
$\phi_M$. Moreover, since $\phi_M$ belongs to the space $V$ we obtain using (\ref{R}) that 
$\psi(k,x) = (R_M^0)^{-1}\varphi_M$ satisfies axiomatics
(\ref{1}),(\ref{2}) in $x$ as well as in $k$ according to the theorem 2.3.
So, we
proved that $\psi(k,x)$ defined by the formula (\ref{formula}) is
the BA function associated to a configuration ${\cal A}$.

\noindent {\it Remark.} The remarkable formula (\ref{formula}) for $\psi$ was discovered by
Yu. Berest (\cite{B2}), who proved that if $\psi$ does exist then it should have
the form (\ref{formula}).

\section* {4. Analysis of the locus equations and locus configurations.}

The next step would be to classify all the solutions of the locus equations 
({\it locus configurations}). Unfortunately, this problem seems to be very difficult. In this section we present  some results in this direction and all the known examples.

\noindent {\bf 4.1. Coxeter systems.}

The most natural examples of the locus configurations are given by the mirrors of
the Coxeter groups. Recall that a Coxeter group $W$ is by definition a finite
group generated by some orthogonal reflections
$s_\alpha(x)=x-\frac{2(\alpha,x)}{(\alpha,\alpha)}\alpha$ with respect to
hyperplanes
in ${\bf R}^n$ (see \cite{Burb}).
If we consider all the reflections from the Coxeter group $W$, then the set
${\mathfrak {A}}$ of the corresponding hyperplanes $(\alpha,x)=0$ will be invariant
under the action of $W$. The configuration ${\mathfrak {A}}$ of these hyperplanes
with arbitrary
$W$-invariant multiplicities $m_\alpha \in {\bf Z}_+$ gives an example of
locus configuration.
This fact follows immediately from the symmetry of the corresponding
potential $u(x)$
with respect to any reflection $s_\alpha, \alpha \in {\cal A}$.

In this case the Schr\"odinger operator $L$ is the quantum Hamiltonian of
the generalised
Calogero -- Moser system (see \cite{Calogero}, \cite{OP}). The existence of
the BA function for
the root system of type $A_n$ with $m_\alpha=1$ was proved in \cite{ChV1},
where some explicit
formula for $\psi$ has been found. For the general Coxeter system it was
done in
\cite{VSCh}, using the Heckman's formula \cite{H} for the so-called shift
operators in terms of
the Dunkl operators \cite{Dunkl}. Notice that our approach gives a new
proof of this result.

\noindent {\it Remark.} In principle, one may try to extend these examples to the
complex case, by considering a finite group generated by orthogonal
reflections in complex Euclidean space.
However, it is known (see e.g. \cite{Cohen}) that all such groups are
nothing but the
complexified Coxeter groups.

\noindent {\bf 4.2. Deformed root systems.}

The first non-Coxeter locus configuration $A_n(m)$ was introduced in
\cite{VFCh}. It consists of
the following vectors in ${\bf R}^{n+1}$: $e_i - e_j$ with multiplicity $m$
($1\le i <j\le n$)
and $e_i - \sqrt{m}e_{n+1}$ with multiplicity $1$ ($i=1,\ldots, n$). Notice
that for $m=1$ we have
the root system $A_n$. We can allow the parameter $m$ to be
negative simply considering the vectors $e_i-e_j$ with the
multiplicity $-1-m$ in that case (then, of course, we will have a complex configuration in ${\bf C}^{n+1}$).

Corresponding Schr\"odinger operator has
the form:

\begin{equation}
\label{AA}
L= - \Delta + \sum_{i<j}^{n}\frac {2m(m+1)}{(x_i-x_j)^2} +
\sum_{i=1}^{n}\frac{2(m+1)}{(x_i-\sqrt mx_{n+1})^2}.
\end{equation}

In the simplest nontrivial case $n=2$ we have the following configuration
(see figure 1).

%
%
\begin{center}
\includegraphics{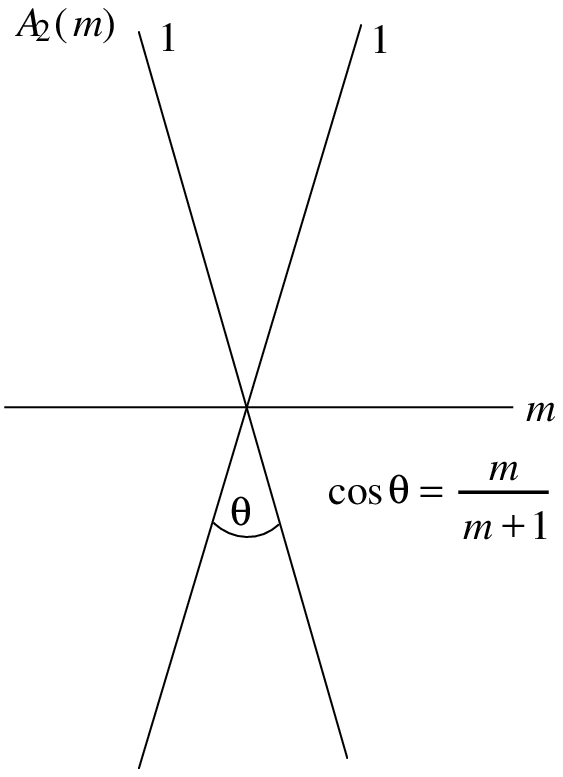}

Figure 1
\end{center}

The next example is related to the root system of $C_n$ - type. Let's consider
the following set of vectors in
${\bf R}^{n+1}$:

\begin{equation}
\label{C}
C_{n+1}(m,l) =
\left\{
\begin{array}{ll}
e_i\pm e_j &  {\rm with \,\, multiplicity \,\,}   k\\
2e_i & {\rm with \,\, multiplicity \,\,}   m\\
2\sqrt{k}e_{n+1} & {\rm with \,\, multiplicity \,\,}  l\\
e_i\pm \sqrt{k}e_{n+1} & {\rm with \,\, multiplicity \,\,}  1\\
\end{array}
\right.
\nonumber
\end{equation}
where $l$ and $m$ are integer parameters such that $k=\frac{2m+1}{2l+1}\in
{\bf Z}$, $1\le i<j\le n$.
In the case of $C_2 (m,l)$ - system the parameters $m,l$ can be arbitrary
integers, the corresponding
quantum problem was considered in \cite{VFCh, ChFV2}. The corresponding
configuration has the form shown on the figure 2.

\begin{center}
\includegraphics{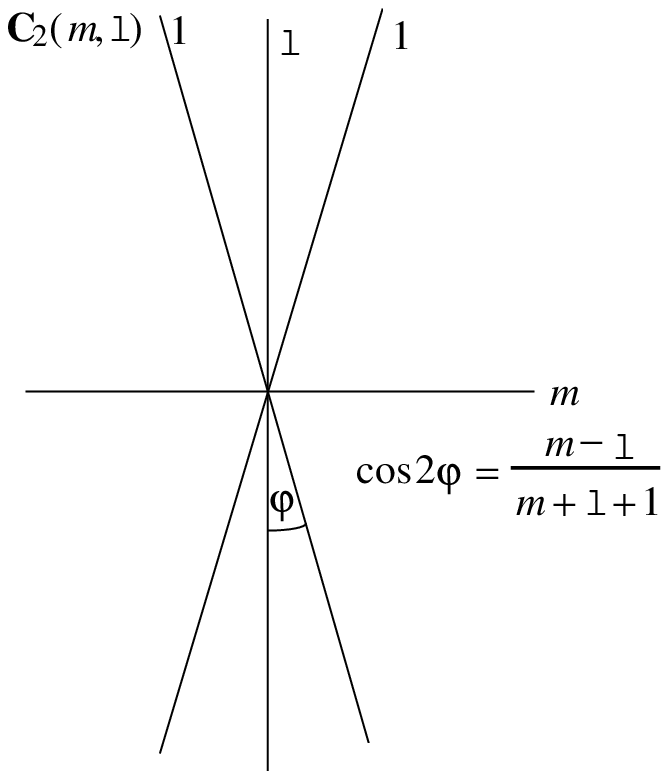}

Figure 2
\end{center}
%
%

For $n>1$ the corresponding Schr\"odinger operator has the
form:
\begin{eqnarray}
L & = & -\Delta_{n+1} +  \sum_{i<j}^n
\frac{4k(k+1)(x_i^2 + x_j^2)}{(x_i^2 - x_j^2)^2} + \sum_{i=1}^n
\frac{m(m+1)}{x_i^2 }+\nonumber\\ &&\label{CC}\\ &&
\frac{l(l+1)}{x_{n+1}^2} + \sum_{i=1}^n \frac{4(k+1)(x_i^2 + k
 x_{n+1}^2)}{(x_i^2 - k x_{n+1}^2)^2},\nonumber \end{eqnarray} where
$k=\frac{2m+1}{2l+1}$.  In the case $l=m$ the system $C_{n+1}(m,l)$
coincides with the classical root system $C_{n+1}$ (or $D_{n+1}$ for
$l=m=0$).  Again, as for the $A_n(m)$ system, the parameters $k, l, m$ may
be negative, in that case the corresponding multiplicities in (\ref{C})
should be $-1-k$, $-1-m$ or $-1-l$ respectively.

The simplest way to check the validity of the locus equations for these
configurations is to
use the following important property of the system (\ref{togd}):

{\bf Theorem 4.1.} {\it A configuration ${\mathfrak {A}}$ satisfies locus equations
(\ref{togd})
if and only if each two-dimensional subsystem of ${\cal A}$ gives a locus
configuration.
In other words, for each two-dimensional plane $\pi\subset {\bf C}^n$ the
vectors
$\alpha\in {\cal A} \cap \pi$ with their multiplicities $m_\alpha$ must
satisfy the locus equations.}

\noindent {\it Remark.} Notice the analogy with the similar property of the Coxeter
and root systems.

{\bf Proof.}
Let us denote by $\widehat\beta$ the orthogonal projection of a vector $\beta$
onto the hyperplane $(\alpha,x)=0$, then $(\widehat\beta,x)\equiv (\beta,x)$
on this hyperplane. Let $\pi<\alpha,\gamma>$ denote the two-dimensional
plane spanned by $\alpha$ and $\gamma\ne\alpha$. Then the subsum of
(\ref{togd}) over $\beta\in \pi<\alpha,\gamma>$ becomes proportional to
$(\widehat\gamma,x)^{-2j-1}$ being restricted to the hyperplane
$(\alpha,x)=0$.
All these subsums for different two-dimensional hyperplanes are independent,
so we come to the following equivalent form of (\ref{togd}):

for any two-dimensional plane $\pi\in{ \bf C } ^n$ and for each
$\alpha\in{\cal A}\cap\pi$ and $j=1,\ldots,m_\alpha$
\begin{equation}
\label{togd2}
\sum_{\beta\in{\cal A\cap\pi} \atop {\beta\ne\alpha}}
m_\beta(m_\beta +1)(\beta,\beta)(\alpha,\beta)^{2j-1}(\beta,x)^{-2j-1}
\equiv 0  \mbox{  for  } {(\alpha,x)=0}.
\end{equation}
That gives the statement of the theorem.

If we analyse the configurations $A_n(m)$ and $C_{n+1}(m,l)$ from this
point of view,
we will have in each two-dimensional plane either a usual root system or
one of their
deformations $A_2(m)$ and $C_2(m,l)$. For these two cases the locus
equations can be
checked by direct calculation.

One can see that our configurations $A_n(m)$ and $C_{n+1}(m,l)$ have one
common feature:
they are obtained from Coxeter configurations by adding a special orbit of
the Coxeter group
with multiplicity $1$ (a sort of "one-orbit deformation" of a Coxeter
configuration).
The following result demonstrates that such property is not accidental:
the hyperplanes with large multiplicities always form a Coxeter subsystem.

{\bf Definition.} {\it Let's say that the hyperplane $\Pi_\beta\in{\mathfrak {A}}$ 
has a large multiplicity $m_\beta$
if in each two-dimensional plane containing vector $\beta$ there are no more than
$m_\beta +1$ vectors
from ${\cal A}$ (without taking into account the multiplicities).}

{\bf Theorem 4.2.} {\it The set ${\bf B}\subset{\mathfrak {A}}$ of all hyperplanes with
large multiplicities forms a Coxeter configuration and all other hyperplanes
and their multiplicities are invariant under
the action of this Coxeter group.}

{\bf Proof.} We shall prove that for each $\beta\in{\cal B}$ the
corresponding reflection
$s_\beta$ preserves the set ${\mathfrak {A}}$ together with multiplicities. This
implies, in particular,
that $s_\beta({\bf B})\subset{\bf B}$. To prove the invariance of ${\mathfrak {A}}$ under $s_\beta$
let's consider as in theorem 4.1 an arbitrary two-dimensional plane$\pi$,
which contains $\beta$,
and the corresponding two-dimensional locus equation (\ref{togd2}):
$$
\sum_{\gamma\in{\cal A\cap\pi} \atop {\gamma\ne\beta}}
m_\gamma(m_\gamma
+1)(\gamma,\gamma)(\beta,\gamma)^{2j-1}(\gamma,x)^{-2j-1}|_{(\beta,x)=0}
\equiv 0 ,
$$
where $j=1,\ldots, m_\beta$. Now we look at these equations for generic
fixed $x$ as a linear
system for unknowns $z_\gamma =
m_\gamma(m_\gamma+1)(\gamma,\gamma)(\beta,\gamma)(\gamma,x)^{-3}$
of the form
\begin{equation}
\label{van}
\sum_{\gamma\ne\beta}
z_\gamma\left(\frac{(\beta,\gamma)^2}{(\gamma,x)^2}\right)^{j-1}|_{(\beta,x)=0}
\equiv 0, \quad j=1,\ldots,
m_\beta.
\end{equation}
We need the following elementary lemma:

{\bf Lemma.} {\it If three unit vectors $\beta,\gamma,\gamma' $ belong to
some two-dimensional subspace in
${\bf C}^n$ and
$$
\frac{(\beta,\gamma)^2}{(\gamma,x)^2} = \frac{(\beta,\gamma')^2}{(\gamma',x)^2}
$$
for all $x$ such that $(\beta,x)=0$ then either $\gamma = \pm\gamma'$ or
$s_\beta(\gamma)=\pm\gamma'$.}

Let's regroup the terms in (\ref{van}) into the groups corresponding to
different values of
$\frac{(\beta,\gamma)^2}{(\gamma,x)^2}$.
From the properties of the Vandermond determinant we easily conclude that
the sum of $z_\gamma$
in each group should vanish. On the other hand, using lemma we see that
there are only two terms
in each group, and they correspond to the pairs of vectors $\gamma,
\gamma'$ with $s_\beta(\gamma)
=\pm\gamma'$. Finally, we arrive to the condition
$z_\gamma+z_{\gamma'}|_{(\beta,x)=0}=0$,
which gives $m_\gamma(m_\gamma+1) = m_{\gamma'}(m_{\gamma'}+1)$, i.e.
$m_\gamma=m_{\gamma'}$.

\noindent {\it Remark.} The Schr\"odinger operators (\ref{AA}), (\ref{CC}) remain
integrable
in a usual (Liouville) sense for the general (non-integer) values of the parameters $m,l$:
there exists at least $n= {\rm dim} V$ independent commuting operators $L_1=L,
L_2,\ldots, L_n$.
Indeed, for $A_n(m)$ case ($m$ is integer) it's easy to check that
the polynomials $p_s = k_1^s+k_2^s+\ldots+k_n^s + m^{\frac{s-2}2} k_{n+1}^s
\,(s=1, 2, \ldots)$
satisfy the conditions (\ref{quasi}) and, according to the theorem 1.2,
there exist differential
operators $L_s$ with the highest symbols $p_s$ such that $L_s\psi=p_s\psi$
and therefore $[L_s, L_t]=0$ (see the explicit formula (\ref{bform})).
Since the coefficients of these operators depend on $\sqrt{m}$ in a rational way,
one can define such operators for general $m$. For $s=2$ one has the
Schr\"odinger operator (\ref{AA}), and other $L_s$ give its quantum
integrals.  In the case of $C_{n+1}(m,l)$- system the similar arguments
prove the integrability of the Schr\"odinger operator (\ref{CC}) for the
general $l,m$, and the commuting quantum integrals $L_s$ have the symbols
$p_s = k_1^{2s}+\ldots+k_n^{2s}+q^{s-1}k_{n+1}^{2s} \,(q=\frac{2m+1}{2l+1},
s=1,2,\ldots)$.

\noindent {\bf 4.3. Locus configurations on the plane. }

Yu.Berest and I.Lutsenko \cite{BL}
 in the context of the Huygens' Principle have introduced the
following family of the real potentials $u$ on the real plane.
In the polar coordinates they have a form
\begin{equation}
\label{BL}
u(r,\varphi)=-\frac2{r^2}\frac{\partial^2}{\partial{\varphi^2}}
\log W[\chi_1(\varphi),\ldots,\chi_M(\varphi)],
\end{equation}
where $\chi_j(\varphi) = \cos(k_j \varphi +\theta_j), \, k_M>\ldots>k_1> 0,
\, k_j \in {\bf N},\, \theta_j \in {\bf R}$ and $W[\chi_1,\ldots,\chi_M]$ is
the Wronskian of $\chi_1,\ldots,\chi_M$.

One can consider the natural complexification of the Berest--Lutsenko family
in the following way. The set of all non-isotropic lines in ${\bf C}^2$ is
isomorphic
to the cylinder ${\bf C}^*\simeq {\bf C}P^1\backslash\{0,\infty\}$ and can
be parametrised
by a complex parameter $\varphi ({\rm mod} \pi)$
$$
x \cos\varphi + y \sin\varphi  = 0.
$$
Any configuration corresponds to a finite number of points in  ${\bf C}^*$:
$\varphi_1,\ldots,\varphi_N$ with multiplicities $m_1,\ldots,m_N$. The corresponding
potential has the form
\begin{equation}
\label{trigu}
u = \frac1{r^2}\sum_{j=1}^N \frac{m_j(m_j+1)}{\sin^2(\varphi - \varphi_j)},
\end{equation}
where $r^2 = x^2 + y^2 \in {\bf C}\backslash\{0\}$ and $\varphi({\rm mod} \pi) =
\arctan\frac{y}{x}$.
The complex Berest--Lutsenko potentials given by the formula (\ref{BL})
with the {\it complex}
parameters $\theta_j$, have the form (\ref{trigu}) with $\varphi_j$ being
the roots of the
trigonometric polynomial $W[\varphi]$, their multiplicities are known to
have a "triangular"
form $\frac{m_j(m_j+1)}2$ (see \cite{AMM}).

{\bf Theorem 4.3.} {\it  All the locus configurations on the plane are
determined by the complex
Berest--Lutsenko formula (\ref{BL}).}

{\bf Proof.} First of all the locus equations (\ref{togd}) in this case are
equivalent to the
following one-dimensional locus equations (cf. \cite{AMM}) for the potential
$
v(\varphi) =\sum_{j=1}^N \frac{m_j(m_j+1)}{\sin^2(\varphi - \varphi_j)}
$:
$$
{\left(\frac{d}{d\varphi}\right)}^{2s-1}\left.\left(\sum_{j\ne i}
\frac{m_j(m_j+1)}{\sin^2(\varphi - \varphi_j)}
\right)\right|_{\varphi=\varphi_i} = 0 \quad (i=1,\ldots, N,\,
s=1,2,,\ldots, m_i).
$$
Now we can use the result from \cite{Ch}, which says that in its turn this
is equivalent to the
existence of the differential operator $D$ with $\pi$-periodic
coefficients, intertwining
the operator ${\cal L}=-\frac{d^2}{d\varphi^2}+v(\varphi)$ with ${\cal L}_0
= -\frac{d^2}{d\varphi^2}$:
\begin{equation}
\label{intert}
{\cal L}\circ D = D\circ {\cal L}_0.
\end{equation}
The idea of the proof is close to the one demonstrated in the proof of the
theorem 3.1, and we shall not reproduce it here.

So, the only remaining thing to prove is that the relation (\ref{intert}) implies 
that ${\cal L}$ can be obtained from ${\cal L}_0$ by classical Darboux transformations.
Let's assume that $D$ has the minimal order among all the intertwiners
of ${\cal L}$ and ${\cal L}_0$
and consider its kernel: $V=Ker D$. As it follows from (\ref{intert}) $V$
is invariant under ${\cal L}_0$:
if $D f = 0$ then $D({\cal L}_0 f) = {\cal L} D f = 0$. Due to
$\pi$-periodicity of the coefficients
of $D$, $Ker D$ is also invariant under the shift $T: f(\varphi)\to
f(\varphi+\pi)$.

We would like to show that the spectrum of ${\cal L}_0|_V$ is simple and
has the form \\$(k_1^2, k_2^2,\ldots, k_M^2)$,
where $0<k_1<k_2<\ldots<k_M$ are some integers. Suppose that there exists
an eigenfunction $f\in V$ with
the eigenvalue $\lambda\ne k^2, k\in {\bf Z}$. Since ${\cal L}_0$ commutes
with $T$, we can assume that
$f$ is a Bloch eigenfunction:
$$
\left\{
\begin{array}{rcl}
{\cal L}_0 f = \lambda f\\
T f = \mu f\\
\end{array}
\right.
$$
If $\lambda\ne k^2$ $f$ has to be pure exponent: $f = C e
^{\sqrt{-\lambda}\varphi}$ or
$f = C e ^{-\sqrt{-\lambda}\varphi}$.
Since $D f =0$ the operator $D$ can be factorised as
$$
D=\widetilde D\circ F, \quad F=\frac{d}{d\varphi} - \frac{f'}{f},
$$
where $\widetilde D$ is a $\pi$-periodic differential operator of the order
one less than
$D$ (see e.g. \cite{Ince}). When $f = C e ^{\pm\sqrt{-\lambda}\varphi}$ we have
$F = \frac{d}{d\varphi}\pm\sqrt{-\lambda}$ and ${\cal L}\circ\widetilde
D\circ F = \widetilde D
\circ F\circ {\cal L}_0 = \widetilde D\circ {\cal L}_0 \circ F$.
Thus $ {\cal L}\circ\widetilde D = \widetilde D
\circ {\cal L}_0$, so $\widetilde D$ is also an intertwiner with the order
one less than the order of $D$.

Thus the spectrum of ${\cal L}_0|_V$ consists only of the squares of integers:
$\lambda = k^2, k\in{\bf Z}$. The same arguments show that $\lambda\ne 0$.
So we have only
to prove that the spectrum is simple. First of all there could be only one
eigenfunction, corresponding
to a given $\lambda = k^2$. Indeed, otherwise $Ker D$ contain the whole
$Ker ({\cal L}_0 -\lambda)$
and therefore $D$ can be factorised $D=D_1\circ({\cal L}_0 - \lambda)$
with $D_1$ being another intertwiner of less order.
Suppose that ${\cal L}_0$ has a Jordan block with $\lambda = k^2$. Consider
the Jordan basis
$f_0, f_1, \ldots: ({\cal L}_0 - \lambda) f_0 =0, ({\cal L}_0 - \lambda)
f_1 =f_0, \ldots$.
Since $f_0$ can not be pure exponent (see above), $f_0 = A\cos (k\varphi +
\theta_0)$, then
$f_1 = \frac{A\varphi}{2 k}\sin (k\varphi + \theta_0) +  B\cos (k\varphi +
\theta_1)$. Now from the
invariance of $Ker D$ under the shift $T$ we conclude that $\frac{A\pi}{2
k}\sin (k\varphi + \theta_0)$
also belongs to $Ker D$. Together with $f_0$ the last function generates
$Ker({\cal L}_0 - \lambda)$,
which leads to factorisation $D = D_1\circ({\cal L}_0 - \lambda)$ and
reducibility of $D$.

Thus we have proven that $Ker D$ is generated by the functions
$\chi_1,\ldots,\chi_n$ of the form
 $\chi_j= \cos(k_j \varphi +\theta_j)$. The general formula (see e.g.
\cite{Crum}) from the theory
of Darboux transformations says that $u=-2\frac{d^2}{d \varphi^2}\log
W[\chi_1,\ldots,\chi_n]$.
The theorem is proven.

We should mention that although the formula (\ref{BL}) is explicit, it is not
so easy to extract the
geometric information about the locus configurations. For example, it is
not clear how to prove the
following theorem using this formula.

It is very easy to show that all two-line locus configurations consist of two 
perpendicular lines with arbitrary multiplicities.
Let's consider the first non-trivial case of three lines $(\alpha,x)=0, \,
(\beta,x)=0$ and $(\gamma,x)=0,\, x\in {\bf C}^2$
with arbitrary multiplicities $m_\alpha, m_\beta, m_\gamma \in {\bf Z}_+$,
and ask when they form a locus
configuration. Modulo the natural rotational equivalence we have the following classification.

{\bf Theorem 4.4.} {\it All the three lines locus configurations are listed
below:\\
1) the Coxeter ${A_2}$ configuration with multiplicities $(m,m,m)$;\\
2) the deformed $A_2(m)$ configuration (\ref{AA}) with multiplicities
$(1,1,m)$ when $m$ is positive and $(1,1,-m-1)$ when $m$ is negative;\\
3) the three lines complex Berest-Lutsenko configurations, which can be
parametrised in
this case as:
$$ \alpha = (1,a), \beta = (1,b), \gamma = (0,1): a^2 - ab + b^2 + 1 = 0, $$
 the multiplicities are $(1,1,1)$.}

{\bf Proof.} Let ${\mathfrak {A}}$ be an arbitrary three lines locus
configuration. Let us
consider the first case when ${\mathfrak {A}}$ has at least two vectors with
multiplicities greater
than 1. Then the theorem 4.2 states that ${\mathfrak {A}}$ has to be a Coxeter
$A_2$ - system.
Now let us suppose that there is only one vector $\gamma=(0,1)$ with
multiplicity $m>1$.
The theorem 4.2 states that other two vectors have to be symmetric with
respect to the
vector $\gamma$, so we may fix the normalisation $\alpha =
(1,\lambda),\,\beta=(1,-\lambda)$.
The locus equation (\ref{togd}) for $\alpha$ has a form:
$$
\frac{2(1+\lambda^2)(1-\lambda^2)}{(x-\lambda y)^3} +
\frac{m(m+1)\lambda}{y^3} =0
\,\,\,\mbox{    if   }
\,\,\, x+\lambda y= 0.
$$
From that it immediately follows that $\lambda$ can take only the
following values:
$\lambda = \pm\frac1{\sqrt{2 m+1}}, \pm\frac{i}{\sqrt{2 m+1}}$, and it is
easy to check that ${\mathfrak {A}}$
is equivalent to the system $A_2(m)$ or $A_2(-m-1)$. The last case we have
to consider is
the case when all the three vectors $\alpha = (1,a), \beta = (1,b), \gamma
= (0,1)$ have multiplicity 1.
The locus equation (\ref{togd}) for vector $\gamma$ takes the form
$$
\frac{2 a(a^2+1)}{(x+a y)^3} + \frac{2 b(b^2+1)}{(x+b y)^3} =0
\,\,\,\mbox{     if   }\,\,\, y= 0
$$
or
$$ (a+b)(a^2+b^2-ab+1)=0.$$
The locus equations (\ref{togd}) corresponding to $\alpha$ and $\beta$ can
be written as follows:
$$
\left\{
\begin{array}{rcl}
(1+a^2)(1+ab)+b(a-b)^3=0\\
(1+b^2)(1+ab)+a(b-a)^3=0\\
\end{array}
\right.
$$
In the case $a+b=0$ this system of equations is fullfilled if and only if
$a^4 =\frac19$, which implies
that ${\mathfrak {A}}$ is either Coxeter system $A_2$ or deformed system
$A_2(-2)$. In the case
$a^2+b^2-ab+1=0$ the above system holds automatically without any
additional restrictions.
Thus, the theorem is proven.

\noindent {\it Remark 1}. We should mention that some of the configurations 3) contain an isotropic line
($a=\pm i, b=0$ or $a=0, b=\pm i$) and therefore actually reduce to the two-line configurations.
Notice also that when $a= i/\sqrt{3} = -b$ we have $A_2(-2)$ configuration.

\noindent{\it Remark 2}. It can be checked that for the configurations 3) from the theorem
4.4 the function $\phi$ with the properties (\ref{0.1}-\ref{0.2}) doesn't exist. This demonstrates that the converse for the statement of the Corollary 2.7 is not true.

Notice that from this result it follows that the locus of $n$ lines is
non-empty only for the special sets of multiplicities.
Moreover, if the locus configuration is real than the set of multiplicities
determines it uniquely up to rotation
due to the following result.

{\bf Theorem 4.5.} {\it There exists no more than one locus configuration
in ${\bf R}^2$ with given cyclically ordered set of
multiplicities.}

{\bf Proof.} Let ${\cal A} = \{\alpha_1,\ldots,\alpha_N\}$ be such a
configuration for given set of multiplicities
$\{m_1,\ldots,m_N\}$, and let us fix normalisation $\alpha_i =
(-\sin\varphi_i, \cos\varphi_i),\,
0\le\varphi_1<\varphi_2<\ldots<\varphi_N<\pi$. Considering the locus equations,
we have, in particular, that
$$
\sum_{{j=1} \atop {j\ne i}}^N \frac{m_j(m_j+1)\cos(\varphi_j
-\varphi_i)}{\sin^3(\varphi_j - \varphi_i)} = 0 \,\,\,{\rm for}\,\,\,i=1,\ldots , N. 
$$
Let's now introduce the function
$$
U(\varphi_1,\ldots,\varphi_N) =
\sum_{i<j} \frac{m_i(m_i+1)m_j(m_j+1)}{\sin^2(\varphi_i - \varphi_j)}.
$$
We conclude that if $\Phi=(\varphi_1,\ldots,\varphi_N)$ defines a locus
configuration then necessarily
$$
\frac{\partial}{\partial\varphi_i} U(\varphi_1,\ldots, \varphi_N)=0.
$$ Function $U$ being a sum of
convex functions is a convex function in the domain
$0\le\varphi_1<\varphi_2<\ldots<\pi$. Suppose
it has one more extremum in the point
$\widetilde\Phi=(\tilde\varphi_1,\ldots,\tilde\varphi_N)$. Then
$U(\varphi_1,\ldots,\varphi_N)$ should be a constant along the segment
$\Phi + (\widetilde\Phi - \Phi)t,\,0\le t\le 1$,
as well as each function $\frac{m_i(m_i+1)m_j(m_j+1)}{\sin^2(\varphi_i -
\varphi_j)}$.
From that it follows that $\tilde\varphi_i = \varphi_i +\varphi_0$ for some
constant $\varphi_0$ for all $i$.
This means that system $\{\alpha_i\}$ is defined uniquely up to a rotation.

{\bf Corollary 4.6.} {\it If all the multiplicities are equal then the only
real configuration
on the plane is Coxeter, i.e. dihedral. }

The consideration of all two-dimensional subsystems implies the following more general result.

{\bf Corollary 4.7} {\it Any real locus configurations in ${\bf R}^n$ with
equal multiplicities must be Coxeter.}

\section* {5. Affine locus.}

In this section we present some results concerning the case, when the
singular set of the potential $u(x)$ of the Schr\"odinger operator is an affine configuration
${\mathfrak {S}}$ of hyperplanes. So, we consider a Schr\"odinger operator $L=-\Delta+u(x)$ with rational potential having second order poles along some 
non-isotropic hyperplanes in ${\bf C}^n$. Let $(\alpha_s,x)+c_s=0$ ($s=1,\ldots , K$) be the equations of these 
hyperplanes. We will suppose also that the potential $u(x)$ decays at infinity, i.e. 
$u(x)\to 0$ while $x\to \infty$ along the rays outside singularities.

Impose now the condition that $L$ has local trivial 
monodromy around its 
singularities. Then the Theorem 2.1 from the section 2 allows us to reformulate this condition as some algebraic conditions on
the arrangement ${\mathfrak {S}}$ of the singular hyperplanes $(\alpha_j,x)+c_j=0$. First of all, it follows that the potential 
$u(x)$ must be of the form
\begin{equation}
\label{6.0}
u(x)=\sum_{j=1}^K \frac{m_j(m_j+1)(\alpha_j,\alpha_j)}
{((\alpha_j,x) + c_j)^2}
\end{equation}
for some integers $m_1,\ldots , m_K$. Then the conditions (\ref{***}) imply that the Schr\"odinger operator with the potential
of the form (\ref{6.0}) has local trivial monodromy around its singularities if and only if 
the following relations are
satisfied:
\begin{equation}
\label{6.1}
\sum_{j\ne i}\frac{
m_j(m_j+1)(\alpha_j,\alpha_j)(\alpha_i,\alpha_j)^{2s-1}}
{((\alpha_j,x) + c_j)^{2s+1}}\equiv 0
\end{equation}
identically on the hyperplane $(\alpha_i,x) + c_i =0$ for all
$i=1,\ldots, K$ and $s=1,\ldots, m_i$.

We will call the relations (\ref{6.1}) as {\it locus equations}. The equations 
(\ref{togd}) from the Section 3 are their particular case, when all the hyperplanes pass through the origin. 
Sometimes we will refer to (\ref{6.1}) and (\ref{togd}) as to affine and linear cases correspondently. 

As it follows from the section 2, the locus equations (\ref{6.1}) are necessary for the existence of a certain
eigenfunction of the corresponding Schr\"odinger operator $L$ (see theorem 2.2).
As well as in the linear case (Section 3) the equations (\ref{6.1}) are sufficient for this. 
The following result has been proven in \cite{Ch}.

{\bf Theorem 5.1}. {\it Let $L=-\Delta+u(x)$ be a Schr\"odinger operator with the potential of the form (\ref{6.0})
which satisfies the affine locus equations (\ref{6.1}). Then $L$ has an eigenfunction $\phi$ of the form 
$\phi(k,x)=P(k,x)exp(k,x)$, where $P$ is polynomial in $k$, $L\phi=-k^2\phi$.}

This eigenfunction (up to a normalization factor ) is given by the Berest's formula analoguous to (\ref{formula}):
\begin{equation}
\label{aformula}
\psi(k,x)= [(-2)^MM!C(k)]^{-1}
(L+k^2)^M[\prod_{j=1}^{K} \left( ({\alpha}_j , x) + c_j\right) ^{m_j}exp(k,x)],
\end {equation}
where $M=\sum_{j=1}^{K} m_j $ and $C(k)=\prod_{j=1}^{K}({\alpha _j},k)^{m_j}$. The normalization is chosen in such way that 
$\psi(k,x) = (1 + o(1) )exp(k,x)$ as $k\to \infty$.

 We start the analysis of the affine locus equations and their solutions (locus configurations) 
 from the one-dimensional case. 
 
 \noindent {\bf 5.1. One-dimensional case.}
   
 In this case we have a configuration of $K$ points
$z_1,\ldots, z_K$ with multiplicities \\$m_1,\ldots, m_K$ on the
complex plane and the potential
$$
u(z)=\sum_{j=1}^K\frac{m_j(m_j+1)}{(z-z_j)^2}.
$$
The locus equations in this case (for $m_j = 1$) have been introduced in the paper
by Airault, McKean and Moser \cite{AMM}. Duistermaat and Gr\"unbaum \cite{DG}
obtained them for the general multiplicities and proved that they are equivalent to the existence of the
differential operator $D$ with rational coefficients, intertwining
$L=-\frac{d^2}{dz^2}+ u(z)$ and $L_0=-\frac{d^2}{dz^2}$:
$$
L\circ D = D\circ L_0.
$$
All such operators $L$ are the results of the classical Darboux
transformations applied to $L_0$, so the potential $u(z)$ can be given in
this case in terms of the Wronskians by the well-known explicit formula:
$$
u(z)=-2\frac{d^2}{dz^2}\log W[\chi_1,\ldots,\chi_m],
$$
where the polynomials $\chi_1,\ldots,\chi_m$ are defined by the recurrent
relations $\chi_1''=0, \chi_2''=\chi_1,\ldots,\chi_m''=\chi_{m-1}$
(see Burchnall--Chaundy \cite{BC}, Adler--Moser \cite{AM}).
The Wronskian is a polynomial $P_m(z,c_1,\ldots,c_m)$ with the
coefficients depending on the additional integration constants
$c_1,\ldots,c_m$ (see \cite{AM} for the details).

Thus, the locus in the one-dimensional case is a union of the rational
algebraic varieties of the dimensions $m=1,2,3,\ldots$, parametrised by
$c_1,\ldots,c_m$, and the locus configurations are simply the roots
of the corresponding
Schur polynomials $P_m(z,c_1,\ldots,c_m)$.
The solution $\psi$ of the corresponding Schr\"odinger equation
$-\psi''+u(z)\psi = -\lambda ^2\psi$ has a form
\begin{equation}
\label{6.2}
\psi = \left(1+\sum_{i=1}^m a_i(z)\lambda ^{-i}\right)e^{\lambda z}.
\end{equation}
This is a degenerate rational case of the hyperelliptic BA function,
corresponding to a general finite-gap operator \cite{DMN}. These
rational BA functions $\psi$ (\ref{6.2}) are characterized by the following
properties in the spectral parameter (cf. \cite{SW}). Let $\xi_1,\ldots,\xi_m$ be arbitrary
parameters, $\psi_s$ be the Laurent coefficients of $\psi$ at $\lambda =0$:
$\psi = \sum_{s=-m}^{+\infty} \lambda ^s\psi_s(z)$. Impose the following $m$
linear conditions on the coefficients $\psi_{-m},\ldots,\psi_{m-1}$:
\begin{equation}
\label{6.3}
\left\{
\begin{array}{l}
\psi_{m-1}+\sum\limits_{s=1}^m \xi_s\psi_{m-2s} = 0 \\
\psi_{m-3}+\sum\limits_{s=1}^{m-1} \xi_s\psi_{m-2s-2} = 0 \\
\psi_{m-5}+\sum\limits_{s=1}^{m-2} \xi_s\psi_{m-2s-4} = 0 \\
\ldots\\
\psi_{-m+1}+ \xi_1\psi_{-m} = 0
\end{array}
\right.
\end{equation}
They are equivalent to a non-degenerate system for $m$ unknown functions
$a_i(z)$ and determine $\psi$ of the form (\ref{6.2}) uniquely. The usual
arguments \cite{K}, \cite{SW} show that such a function satisfies
the Schr\"odinger equation $-\psi''+u(z)\psi = -\lambda ^2\psi$ with the
rational potential
\begin{equation}
\label{6.40}
u(z)=2a_1'(z).
\end{equation}
Notice that for given $\xi_1,\ldots,\xi_m$ the system (\ref{6.3})
determines a $m$-dimensional linear subspace $V(\xi_1,\ldots,\xi_m)$
in ${\bf C}^{2m}$ and therefore corresponds to a point of the
Grassmannian $Gr(m,2m)$. It is more convinient to identify the system of conditions
(\ref{6.3}) with a point of some infinite-dimensional Grassmannian $Gr_0^{(2)}$ (see \cite{SW} for the details). 
Namely, let's consider the linear space ${\bf C}[[\lambda]]$ of formal series in $\lambda$,
and let $W$ be a subspace of ${\bf C}[[\lambda ]]$ with the 
following properties:\\
1) $\lambda ^m {\bf C}[\lambda ] \subset W \subset \lambda ^{-m} {\bf C}[\lambda]$
where ${\bf C}[\lambda ]$ is the space of polynomials and both inclusions have the same codimension $m$;\\
2) $\lambda ^2 W \subset W$.\\
We will suppose that the number $m=m(W)$ in 1) cannot be reduced. The set of all such subspaces for
$m=0,1,2,\ldots $ we will denote as $Gr_0^{(2)}$ following \cite{SW}.

It is easy to see that the subspace of ${\bf C}[[\lambda ]]$ consisting of all 
Laurent series $\psi = \sum_{s=-m}^{+\infty} \lambda ^s\psi_s$ which satisfy the conditions
(\ref{6.3}) represent nothing but a general point of $Gr_0^{(2)}$.
In these notations the one-dimensional BA function corresponding to $W$ is the unique 
element $\psi_W$ of the form (\ref{6.2}) which Laurent expansion at $\lambda =0$ belongs
to $W$ for each $z$. We will denote by $u_W$ the corresponding potential (\ref{6.40}).

These considerations suggest
the following extension of the axiomatics (\ref{1}-\ref{2}) of the multidimensional BA function.

\noindent {\bf 5.2. Equipped configurations and BA functions.}

Let ${\cal A}$ be again a finite set of non-collinear vectors in
${\bf C}^n$. We will prescribe to each vector $\alpha\in{\cal A}$ a
subspace $W^{(\alpha)}\in Gr_0^{(2)}$, and denote the corresponding integer 
$m(W^{(\alpha)})$ as $m_{\alpha}$. 
We will call the
corresponding set of hyperplanes $\Pi_\alpha:(\alpha,k)=0$ with the
prescribed subspaces $W^{(\alpha)}$ as {\it equipped configuration} ${\mathfrak {A}}$. 

{\bf Definition}. {\it For a given equipped configuration ${\mathfrak {A}}$ the function
$\psi(k,x)$ is called the Baker-Akhiezer function if it satisfies the following two conditions:\\
1) $\psi$ has the form
\begin{equation}
\label{6.4}
\psi=\frac{P(k,x)}{A(k)}e^{(k,x)},
\end{equation}
where $A(k)=\prod_{\alpha\in {\cal A}} (\alpha,k)^{m_\alpha}$, $P$ is a
polynomial in $k$ with the highest term $A(k)$;\\
2) for each $\alpha\in {\cal A}$ the Laurent expansion of $\psi$ in $k$
in $\alpha$-direction calculated at any point of the hyperplane $\Pi_\alpha$ must belong to $W^{(\alpha)}$.}

Here by the Laurent expansion of a meromorphic function $F(k)$ in $\alpha$-direction at a point $k_0$ we mean the Laurent expansion of the function 
$f(\lambda)= F(k_0 + \lambda \alpha)$ at $\lambda = 0$.

If for each subspace $W^{(\alpha)}$ 
the corresponding parameters $\xi$ in (\ref{6.3}) are zeros, our definition  reduces to the definition
of the BA function from the Section 1. Now we will prove the analogues
of the theorems 1.1, 1.2 for a general equipped configuration.

{\bf Theorem 5.2.} {\it If for a given equipped configuration ${\mathfrak {A}}$ there
exists BA function $\psi$ then it is unique and satisfies the
Schr\"odinger equation
\begin{equation}
\label{6.5}
\left( -\Delta +\sum_{\alpha\in{\cal A}} 
(\alpha, \alpha)u_\alpha((\alpha,x))\right)\psi =-k^2\psi,
\end{equation}
where $u_\alpha(z)= u_{W^{(\alpha)}}(z)$ are the one-dimensional potentials, corresponding to
the subspaces $W^{(\alpha)}$.}

{\bf Theorem 5.3.} {\it Let ${\cal R}$ be the ring of polynomials $f(k)$
with the following properties:

for each $\alpha \in {\cal A}$ and any point $k_0\in \Pi_\alpha$ the polynomial
$f_{\alpha, k_0}(\lambda) = f(k_0 + \lambda \alpha)$ preserves the space $W^{(\alpha)}$: $ f_{\alpha, k_0} W^{(\alpha)}\subset W^{(\alpha)}$.

If the Baker-Akhiezer function $\psi ( k, x)$ exists then for any polynomial
$f ( k ) \in { \cal R}$ there exists some differential operator
$L_f ( x,\frac { \partial} { \partial x } )$
such that
$$
L_f \psi ( k, x ) = f ( k)\psi ( k, x ).
$$
All such operators form a commutative ring isomorphic to
the ring $ {\cal R}$.  The Schr\"odinger operator (\ref{6.5})
corresponds to $f(k)=-k^2$.}

The proofs of the theorems above follow in a standard way (cf. \cite{ChV1})
from the following two lemmas.

{\bf Lemma 1}. {\it If some function $\psi$ of the form (\ref{6.4})
(without the restrictions on the highest term of the polynomial $P$)
satisfies the conditions 2 from the definition of the BA function 
then
the highest term in $P$ must be divisible by $A(k)=
\prod_{\alpha\in {\cal A}} (\alpha,k)^{m_\alpha}$.}

{\bf Lemma 2}. {\it The BA function corresponding to an equipped 
configuration ${\mathfrak {A}}$ has the following asymptotic behaviour at infinity:
$$
\psi(k,x) = exp(k,x) \left( 1 +  \sum_{\alpha\in {\cal A}} a_1^{(\alpha)}((\alpha,x))\frac {(\alpha, \alpha)}{(\alpha,k)}+ o(k^{-1})\right)
$$
where $a_1^{(\alpha)}(z)$ are the first coefficients in the
corresponding functions (\ref{6.2}) $\psi_\alpha = \psi_{W^{(\alpha)}}$
and $o(k^{-1})$ means the rational function of k which degree is less than -1.}

To prove the lemmas, let's expand $\psi$ in Laurent series in $(\alpha,k)$
on the hyperplane $(\alpha,k)=0$. For convenience we may suppose that $(\alpha, \alpha) = 1$ and choose
orthonormal basis in $k$ such that $(\alpha,k)=k_1$, the other coordinates
$k_2,\ldots, k_n$ we shall denote by $\tilde k$. Then up to unessential
factor $exp(k_2 x_2 +\ldots +k_n x_n)$
$\psi$-function (\ref{6.4}) takes the form:
\begin{equation}
\label{6.6}
\tilde\psi(k,x) = e^{x_1 k_1}\sum_{s\ge - m_\alpha} k_1^s a_s(\tilde
k,x),
\end{equation}
and the Laurent coefficients $a_s$ are rational functions of
$\tilde k$ with possible singularities at zeros of homogeneous
polynomial $\tilde A(\tilde k) = k_1^{-m} A(k)|_{k_1=0}$.
Since the sum $\sum_{s\ge -m_\alpha} k_1^s a_s$ is the Laurent expansion
for $\frac{P(k,x)}{A(k)}$,
the degrees in $\tilde k$ of its coefficients $a_s$ decrease at
$s\to\infty$ (by definition, ${\rm deg}\frac{p}{q} = {\rm deg}p - {\rm deg}q$).
Now we restrict our attention to the terms $k_1^s a_s$ with the maximal
degree of $a_s$ in $\tilde k$. From the remark above it follows that we
have a finite
number of such terms, and if we extract the highest homogeneous part in
$\tilde k$ in each term, we obtain the following finite expression
\begin{equation}
\label{6.7}
\tilde\psi^0(k_1,x) = e^{x_1 k_1}\sum_{s\ge - m_\alpha} k_1^s
a^0_s(\tilde k,x),
\end{equation}
where $a^0_s$ is the highest term in $a_s$
and all the $a^0_s$ have the same degree in $\tilde k$. It is clear
now that constructed in that way $\tilde\psi^0$ must obey the same
restrictions (\ref{6.3}). This implies, in particular, that the sum
(\ref{6.7}) contains at least one term with $s\ge 0$.
The outcome is that if we expand $P(k,x)$ in the series in $k_1$,
$P=\sum_{j\ge 0} k_1^j p_j(\tilde k)$, and then extract from this
sum the terms with the maximal degree in $\tilde k$, the result must
contain at least one term with $j\ge m_\alpha$. Now let's present $P$
as a sum of homogeneous in $k_1,\ldots, k_n$ components
$P=P_0+P_1+\ldots$, and suppose that the highest term $P_0$ is not
divisible by $k_1^{m_\alpha}$. In this case some other term $P_i$ must
contain  $k_1^{m_\alpha}$, but its degree in $\tilde k$ is clearly less
than the degree of the term coming from $P_0$. This contradiction proves
the lemma 1.

Moreover, in the extreme case when $P_0$ has the form
$P_0=k_1^{m_\alpha}Q_0$ with $Q_0|_{k_1=0}\neq 0$ the reduced
$\psi$-function (\ref{6.7}) up to a factor coincides with the
one-dimensional BA function (\ref{6.2}) $\psi(k_1,x_1)$. It's easy to see
that this factor is simply $Q_0|_{k_1=0}$.

In particular, this implies that the second homogeneous term $P_1$
in $P(k,x)$ for the BA function $\psi$ satisfies the following
condition:
$$
{\left[ k_1^{1-m_\alpha} P_1\right]}_{k_1=0} =
a_1(x_1){\left[ k_1^{-m_\alpha} P_0\right]}_{k_1=0} ,
$$
where $a_1$ is the first coefficient in the corresponding
one-dimensional BA function (\ref{6.2}).
We obtained this formula under assumption that $(\alpha, \alpha) = 1$, in general it looks as follows: 
\begin{equation}
\label{6.8}
{\left[ (\alpha, k)^{1-m_\alpha} P_1\right]}_{(\alpha , k)=0} =
(\alpha, \alpha) a_1((\alpha , x)){\left[ (\alpha, k)^{-m_\alpha} P_0\right]}_{(\alpha, k)=0}.
\end{equation}

Taking into account the restrictions (\ref{6.8}) for all the hyperplanes
$(\alpha,k)=0$, and we obtain that if $P_0 = A(k) =
\prod_{\alpha\in {\cal A}} (\alpha,k)^{m_\alpha}$, then
\begin{equation}
\label{6.9}
P_1 = A(k) \sum_{\alpha\in {\cal A}} a_1^{(\alpha)}((\alpha,x))\frac {(\alpha, \alpha)}{(\alpha,k)} ,
\end{equation}
which proves the lemma 2.

Let's consider now for a given equipped configuration ${\mathfrak {A}}$  the corresponding Schr\"o\-din\-ger operator
(\ref{6.5}). It is clear that the potential has the form (\ref{6.0}). 
The corresponding affine configuration of the hyperplanes ${\mathfrak {S}}$ we will call {\it dual} to the equipped configuration ${\mathfrak {A}}$.  
Suppose that the corresponding BA function does exist, then from the 
theorem 2.2 we conclude that the Schr\"odinger operator (\ref{6.5}) has local trivial monodromy and hence satisfies the locus equations (\ref{6.1}).
In other words, the dual configuration ${\mathfrak {S}}$ must be a locus configuration. 
We believe that the converse is true, that is, {\it each} locus configuration appears in such way 
for appropriate BA function. The part 2 of the Theorem 5.6 below shows that each locus configuration is dual
to some equipped configuration. So, the only problem is to check that for the
function defined by the formula (\ref{aformula}) the properties 2 from the definition of the BA function 
hold. Unfortunately, we couldn't find a proof for this. We can only remark that for all known affine locus configurations it is true. 

\noindent {\bf 5.3. Geometry of affine locus.}

First of all, it is easy to check that the following operations preserve the locus equations and therefore
allow to produce the locus configurations:

1) motions of the complex Euclidean space ${\bf C}^n$;

2) extensions of the configurations in ${\bf C}^n$ to ${\bf C}^m$, $m>n$, induced by an orthogonal 
projection ${\bf C}^m \to {\bf C}^n$;

3) union of two configurations which are orthogonal to each other.

At the moment all known examples of the affine locus configurations can be constructed using these 
operations from one-dimensional affine and multidimensional linear locus configurations.

In particular, this is true for the configurations, corresponding to the operators introduced by Yu.Berest 
and P.Winternitz \cite{BW}. Analysis of these examples, however, reveals one more geometrical way to 
produce the locus configurations.

Let ${\mathfrak {S}}$ be any affine configuration of hyperplanes in ${\bf C}^n$.
Let's imbed ${\bf C}^n$ in ${\bf C}^{n+2}$ in the following way:
$ x = (x_1,\ldots, x_n) \to (x_1,\ldots, x_n, 1, 0)$. For any hyperplane
$\Pi$ in ${\bf C}^n$ let's define the hyperplane $\widetilde\Pi$ in
${\bf C}^{n+2}$ as a linear span of
$\Pi\subset {\bf C}^{n}\subset {\bf C}^{n+2}$ and the
isotropic vector $e=(0,\ldots,0,1,i)$.
If $(\alpha,x)+c=0$ is the equation of $\Pi$ in ${\bf C}^n$ then the
corresponding equation of $\widetilde\Pi$ will be
$(\alpha,x)+c (x_{n+1} + i x_{n+2})=0$.

Corresponding configuration $\widetilde {\mathfrak {S}}$ in ${\bf C}^{n+2}$ we will call
as {\it isotropic projectivisation} of ${\mathfrak {S}}$.

{\bf Theorem 5.4.} {\it The isotropic projectivisation of an affine locus 
configuration ${\mathfrak {S}}$ in ${\bf C}^n$ is a linear locus
configuration $\widetilde {\mathfrak {S}}$ in ${\bf C}^{n+2}$.}

{\bf Proof.} We shall check the first of the locus equations
for $\widetilde {\mathfrak {S}}$, the others can be checked in the same way. So, we need
to prove that on a hyperplane $(\alpha_s,x) + c_s (x_{n+1}+i x_{n+2})=0$
the following identity holds:
$$
\sum_{j\ne s}\frac{
m_j(m_j+1)(\tilde\alpha_j,\tilde\alpha_j)(\tilde\alpha_s,\tilde\alpha_j)}
{((\alpha_j,x) + c_j(x_{n+1}+i x_{n+2}))^3}\equiv 0,
$$
where $\tilde\alpha_j$ denotes
the normal vector of the hyperplane $\widetilde\Pi_j\subset{\bf C}^{n+2}$.
If $\widetilde\Pi_j\subset{\bf C}^{n}$ has the the normal vector
$\alpha_j = (\alpha_j^1,\ldots,\alpha_j^n)$, then $\tilde\alpha_j$ is
the vector $ (\alpha_j^1,\ldots,\alpha_j^n,c_j, i c_j)$.
>From that we immediately see that $(\tilde\alpha_j,\tilde\alpha_j) =
(\alpha_j,\alpha_j)$ and $(\tilde\alpha_s,\tilde\alpha_j) =
(\alpha_s,\alpha_j)$. Now since $\lambda =x_{n+1} + i x_{n+2}\ne 0$ almost
everywhere on the hyperplane $(\alpha_s,x)+c_s(x_{n+1}+i x_{n+2})=0$
we come to the identity
$$
\sum_{j\ne s}\frac{
m_j(m_j+1)(\alpha_j,\alpha_j)(\alpha_s,\alpha_j)}
{((\alpha_j,x) + c_j\lambda)^3}\equiv 0
$$
for $(\alpha_s,x)+c_s\lambda =0$.
But this identity after the rescaling $x\to \lambda x$ takes the form
$$
\sum_{j\ne s}\frac{
m_j(m_j+1)(\alpha_j,\alpha_j)(\alpha_s,\alpha_j)}
{((\alpha_j,x) + c_j)^3}\equiv 0
\qquad \mbox{for}\quad (\alpha_s,x)+c_s=0,
$$
which is exactly the first locus equation for the configuration ${\mathfrak {S}}$.

{\bf Example.} Let ${\mathfrak {S}}$ be a direct sum of three-point
one-dimensional configurations with the corresponding potential
$$
u(x_1,\ldots,x_n)=\sum_{i=1}^n\frac{6x_i^4-12\tau_ix_i}{(x_i^3+\tau_i)^2}.
$$
Then after the isotropic projectivisation we obtain the locus
configuration with the potential of the form (cf. \cite{BW}):
$$
\tilde u(x_1,\ldots,x_{n+2})=\sum_{j=1}^n\frac
{6x_j^4-12\tau_j(x_{n+1}+i x_{n+2})^3 x_j}{(x_j^3+\tau_j(x_{n+1}+i
x_{n+2})^3)^2}.
$$
In order to obtain a more general Berest--Winternitz's potential \cite{BW}
$$
\tilde u(x_1,\ldots,x_{n+2})=\sum_{j=1}^n\frac
{6x_j^4-12\tau_j(x_{n+1}+i x_{n+2} +c_j)^3 x_j}{(x_j^3+\tau_j(x_{n+1}+i
x_{n+2}+c_j)^3)^2}
$$
we should shift the pairwise-orthogonal triples of hyperplanes
$$
x_j+\tau_j^{\frac13}(x_{n+1}+i x_{n+2}) =0
\quad (j=1,\ldots,n)
$$
by $c_j$ in $x_{n+1}$.

\noindent {\it Remark}. The BA function in this example can be obtained 
easily using the following general remark. If $\psi_i=R_i(k,x)exp(k,x)$ \,($i=1,2$) are given by the formula (\ref{aformula}) for two orthogonal locus configurations ${\mathfrak {S}}_1$
and ${\mathfrak {S}}_2$ then the function $\psi=R_1R_2exp(k,x)$ will correspond to the locus configuration ${\mathfrak {S}} = {\mathfrak {S}}_1 \bigcup {\mathfrak {S}}_2$. This is clear from the structure of the formula (\ref{aformula}).

 Thus, iterating such geometric procedures
one can construct many new affine locus configurations. However, all of
them are degenerate in the following sense. Let $V({\mathfrak {S}})$ be the linear
space of the normals to all the hyperplanes in ${\mathfrak {S}}$. We call ${\mathfrak {S}}$
{\it degenerate} if the restriction of the complex Euclidean form on
$V({\mathfrak {S}})$ is degenerate.

For a degenerate affine configuration one can define the following {\it
isotropic reduction} procedure, which is inverse to the isotropic
projectivisation.

Let $K$ be the kernel of the restriction of the Euclidean form onto
$V({\mathfrak {S}})$. Consider the orthogonal complement $V^\bot$ of $V$ in
${\bf C}^n$ and choose a subspace $L$ such that
$$
V+V^\bot = K\oplus L.
$$
By an {\it isotropic reduction} of the degenerate configuration ${\mathfrak {S}}$
we shall mean the configuration ${\mathfrak {S}}\cap\{a+L\}$, where $\{a+L\}$ is a
shift of $L$ by a generic vector $a\in {\bf C}^n$.

{\bf Theorem 5.5.} {\it An isotropic reduction of a degenerate locus
configuration is a non-degenerate locus configuration.}

The proof is similar to the case of isotropic projectivisation.

These results may be interpreted in two ways. First, we can say that any affine locus configuration is a result of the isotropic reduction of some (degenerate) linear configuration. So, the classification problem for affine locus configurations reduces to the linear case. On the other hand, as we have shown,  to classify all locus configurations it is
sufficient to consider non-degenerate configurations only. Moreover, we can consider irreducible configurations only, i.e. exclude the unions of orthogonal subconfigurations. At the moment
all the known non-degenerate irreducible locus configurations are linear or one-dimensional.
It may well be the only possible
examples.

The following general result clarifies the geometrical structure of affine
locus configurations.

{\bf Theorem 5.6.} {\it Any affine locus configuration ${\mathfrak {S}}$ has the
following properties:

(1) for each point $x_0\in {\bf C}^n$ the subset ${\mathfrak {S}}_{x_0}\subseteq
{\mathfrak {S}}$ of the hyperplanes passing through $x_0$ form a linear locus
configuration;

(2) for each hyperplane $\Pi\in\Sigma$ the subset ${\mathfrak {S}}(\Pi)
\subseteq {\cal A}$ of the hyperplanes parallel to $\Pi$ forms an extended
one-dimensional locus configuration.

Conversely, any affine configuration with properties (1), (2) belongs to
the locus.}

{\bf Proof.} (1) Let's consider the locus equations for some
hyperplane $\Pi_i :\,(\alpha_i,x)+c_i=0$ passing through ${x_0}$:
\begin{equation}
\label{afloc}
\sum_{j\ne i}
\frac{m_j(m_j
+1)(\alpha_j,\alpha_j)(\alpha_i,\alpha_j)^{2s-1}}
{((\alpha_j,x)+c_j)^{2s+1}} \equiv 0 \,\,\mbox{  for  }\,\, x\in \Pi_i,
\end{equation}
$s=1,\ldots,m_i$. 

Now take $x=x_0+y$, then $x\in \Pi_i$ iff $(\alpha_i,y)=0$ and we have the following relation:
$$
\sum_{{j:x_0\in \Pi_j}\atop{j\ne i}}
\frac{m_j(m_j +1)(\alpha_j,\alpha_j)(\alpha_i,\alpha_j)^{2s-1}}
{(\alpha_j,y)^{2s+1}} +
\sum_{k:x_0 \notin \Pi_k}
\frac{m_k(m_k +1)(\alpha_k,\alpha_k)(\alpha_i,\alpha_k)^{2s-1}}
{((\alpha_k,x_0)+c_k+(\alpha_k,y))^{2s+1}}
\equiv 0
$$
for all $y$ such that $(\alpha_i,y)=0$. Since the second sum is
regular at $y=0$, the first sum should vanish
on the hyperplane $(\alpha_i,y)=0$. Thus, we obtain linear locus
equation for the configuration ${\mathfrak {S}}_{x_0}$.

(2) To prove the second property, let's divide all the hyperplanes
which are non-parallel to $\Pi$ into the subgroups in the following way:
$\Pi'$ and $\Pi''$ belong to the same group if and only if their
intersection is contained in $\Pi$. Then
in each group the sum of the corresponding terms in (\ref{afloc})
should vanish due to the property (1). The remaining terms is exactly
the locus equation for the set of parallel planes ${\mathfrak {S}}(\Pi)$.

The converse statement now is clear.

We conclude this section by some negative results about locus configurations in  ${\bf R}^n$.

{\bf Theorem 5.7.} {\it For any locus configuration in the real plane
there exists a point all the lines pass through.}

{\bf Proof.} First we note that parallel lines cannot appear in
locus configurations in ${\bf R}^2$. Indeed, the subset of parallel
lines according to the previous theorem must give a real solution
for the one-dimensional locus equations, which is impossible.

Now let's fix some terminology: by vertices we will mean the intersection
points for the lines from the configuration and by a ray -- any ray from
the configuration with the origin at some vertex (some rays may contain
other vertices). Let's choose an orientation on the plane. This allows us
to determine the oriented angle $\varphi(l_1,l_2)$ between the ordered
pair of rays$l_1, l_2$, which varies from $-\pi$ to $\pi$. We need the
following property of the locus configurations in ${\bf R}^2$:

{\bf Lemma 1.} {\it For each ray $l_1$ from the locus configuration in
${\bf R}^2$ there exists another ray $l_2$ with the same vertex and
acute angle between $l_1$ and $l_2$:
$$
  0<\varphi(l_1,l_2)\le\frac\pi2.
$$
Similarly, there exists a ray $l_3$ with the same vertex such that
$-\frac\pi2\le\varphi(l_1,l_3)<0$.}

Proof of the lemma follows from the linear locus equations
(\ref{togd}) for the lines passing through a given vertex: it's clear
that the sign of each term in it depends only on the sign of the
cotangent of the oriented angle between $\alpha$ and $\beta$.

{\bf Lemma 2.} {\it Let $l_1$ and $l_2$ be chosen as in Lemma 1. 
Then if $l_1$ contains another vertex of the configuration,
the same is true for $l_2$.}

The proof follows from simple geometrical considerations.

Let's consider now any vertex and all the rays
of our configuration outgoing from this vertex. As it easily follows from
the lemmas we have only two possibilities:\\
1) there are no other vertices on these rays or\\
2) there is at least one more vertex on each ray.\\
Since we have a finite number of vertices, we obtain immediately that
our configuration has only one vertex. Theorem is proven.

The same is probably true in ${\bf R}^n$ but at the moment we can prove this
only in the special case when all the multiplicities are equal.

{\bf Theorem 5.8.} {\it Any affine locus configuration in ${\bf R}^n$ with
equal multiplicities is a linear Coxeter configuration.}

{\bf Proof.} It's sufficient to prove that the configuration must be
symmetric with respect to each its hyperplane. Since the parallel
hyperplanes cannot appear in a real locus configuration, the statement
follows from the theorem 5.6 and the corollary 4.7.

\section* {6. Locus configurations and Huygens' Principle.}

Let us consider a linear hyperbolic equation
\begin{equation}
\label{HYP}
{\cal L}\varphi(x) = 0, \,\,\, {\cal L} = \Box_{N+1} + u(x),
\end{equation}
where $\Box_{N+1}$ is the D'Alembert operator, 
$\Box_{N+1} = \frac { \partial^2} { \partial x_0^2} - \frac { \partial^2} {
\partial x_1^2}-\ldots-\frac { \partial^2} { \partial x_N^2 }$.

We say after J.Hadamard \cite{Had} that it  satisfies {\it Huygens'
Principle} (HP) if its
fundamental solution is located on the characteristic conoid, i.e.
this solution vanishes
in the conoid's complement.

Hadamard found some criterion for HP to be satisfied
in terms of the so-called Ha\-da\-mard's coefficients $U_\nu ( x, \xi ).$
They are uniquely determined by the following system of equations
\begin{equation}
\label{Had}
\sum_ { i=0}^N ( x_i-\xi_i ) \frac { \partial U_\nu} { \partial x_i } + \nu
U_\nu = -\frac12 { \cal L} ( U_ { \nu -1 } )
\end{equation}
and the conditions that $U_0 ( x, \xi ) \equiv 1$ and $U_\nu (x, \xi)$ are
regular at $x = \xi$.
These coefficients are symmetric with respect to $x$ and $\xi$:
$
U_\nu ( x, \xi ) =U_\nu ( \xi, x )
$
(for the details see the book \cite{Gun}).

Hadamard proved that the equation (\ref{HYP}) satisfies Huygens' Principle
if and only if $N$ is odd and $U_\nu |_\Gamma = 0$ for $\nu\ge\frac{N-1}{2}$,
where $\Gamma=\{(x,\xi): (x_0-\xi_0)^2 - \sum_ { i=1}^N ( x_i-\xi_i )^2 =
0\}$ is
the characteristic conoid.
For the case when potential $u$ (and, as a corollary, all the Hadamard's
coefficients
$U_\nu$) does not depend on at least one of the coordinates (say, $x_0$), the Hadamard's criterion is equivalent to
the condition
$U_{\frac{N-1}2}\equiv 0$.

 We consider the Hadamard's problem of description of all huygensian equations of the form:
\begin{equation}
\label{HYPP}
(\Box_{N+1} + u(x_1,\ldots,x_N))\varphi = 0
\end{equation}
In fact, in our case for any locus configuration in ${\bf C}^n$ the
corresponding potential
will depend only on the first $n$ coordinates : $u=u(x_1,\ldots, x_n),\, n\le N$.

It turns out that huygensian equations of the form (\ref{HYPP}) are closely
related to the locus configurations. For the linear locus configuration in  ${\bf C}^n$ the corresponding potential 
\begin{equation}
\label{6.00}
u(x) = \sum_{\alpha\in {\cal A}}\frac
{m_\alpha (m_\alpha +1) (\alpha,\alpha)}{(\alpha,x)^{2}}.
\end{equation}
is homogeneous of degree --2.

{\bf Theorem 6.1.} {\it For any real potential $u(x_1,\ldots, x_n)$ related
to a linear locus
configuration the hyperbolic equation (\ref{HYPP})
satisfies HP if
$N$ is odd and $N\ge 2\sum_{\alpha\in {\cal A}} m_\alpha +3$.
In that case the fundamental solution can be expressed via BA function.

Conversely, if the hyperbolic equation (\ref{HYPP})
with
homogeneous potential $u(x)$: $u(\lambda x) = \lambda^{-2}u(x)$ satisfies  HP
and all the Hadamard's coefficients are rational functions, then the potential $u(x)$ must have the form (\ref{6.00}) for some linear locus configuration.}

{\bf Proof.}  The  proof of the first statement repeats the arguments of
the paper \cite{BV},
where this result has been proven in the Coxeter case. It is based on the
following relation
between BA function and Hadamard's coefficients.
If we have the Baker-Akhiezer function $\psi$
of the form (\ref{1}), we can present it in the form
\begin{equation}
\label{Fi}
\psi(\xi,x) = (U_0(\xi,x)+U_1(\xi,x)+\ldots+U_M(\xi,x)) e^{(\xi,x)},
\end{equation}
where $U_0=1, U_\nu (x,\xi)$ is homogeneous of degree $-\nu$ in $\xi$,
$M=deg A(k)=\sum_{\alpha\in{\cal A}} m_\alpha$. Since $\psi$ is symmetric in
$\xi$ and $x$ (theorem 2.3), $U_\nu$ has the same degree in $x$. From the
Schr\"odinger equation (\ref{Seq}) for $\psi$, $L\psi=-\xi^2\psi$, $L=-\Delta + u(x)$, we obtain:
$$
-2\sum_{i=1}^n \xi_i\frac{\partial}{\partial x_i} U_\nu + L[U_{\nu-1}] = 0
\quad (\nu=1,\ldots, M+1 \mbox{   with   } U_{M+1}=0) .
$$
Since $U_\nu$ are homogeneous in $x$ this implies the relations (\ref{Had}), so $U_\nu$ coincide with the Hadamard's coefficients. 
Now since $U_{M+1}=0$ the Hadamard's criterion guarantees HP if $N\ge 2M+3$.   
Notice that it gives also the explicit formula for the Hadamard's coefficients and the fundamental solution for (\ref{HYP}) (see for the details \cite{BV}).

Conversely,
from the chain (\ref{Had}) for the Hadamard's coefficients $U_\nu(x,\xi)$
for the homogeneous potential $u$ it follows that $U_\nu$ are also
homogeneous in $x$ (and, therefore, in $\xi$):
$$
U_\nu(\lambda x,\xi)=\lambda^{-\nu} U_\nu(x,\xi) = U_\nu(x, \lambda\xi).
$$
This can be proven by the same calculation as in lemma 1 from \cite{B}, where the case $n=2$ was considered.
Let's now consider the function $\psi$ defined by the formula (\ref{Fi}).
Then,  from the Hadamard chain (\ref{Had}) and homogeneity of
$U_\nu$ it follows in the same way as above that $\psi$ satisfies the Schr\"odinger equation
$$
(-\Delta_N + u(x))\psi = - \xi^2\psi.
$$
Notice that the potential $u(x)$ must be rational since all the Hadamard's coefficients are supposed to be rational. This follows from the first equation of the Hadamard's chain (\ref{Had}). Now using the theorems 2.1 and 2.2 and the fact that $u(x)$ is homogeneous of degree (--2) we conclude that $u(x)$ has the form (\ref{6.00}) for some locus configuration.

\noindent {\it Remark.} In the case when $n=2$ i.e. $u=u(x_1, x_2)$, a stronger result (namely, without the assumption that the Hadamard's coefficients are rational) follows from the results by Yu.Berest and I.Lutsenko \cite{BL}, \cite{B}.

Now let's consider an arbitrary (affine) locus configuration ${\mathfrak {S}}$ such that the corresponding potential $u(x)$ given by the formula (\ref{6.0}) is real for real $x$. This is equivalent to the condition ${\mathfrak {S}} =\bar\Sigma$ where $\bar {\mathfrak {S}}$ is a natural complex conjugation of a configuration
${\mathfrak {S}}$. The following result generalises the Theorem 6.1 for the general (affine) locus configurations.

{\bf Theorem 6.2.} {\it For any affine locus configuration ${\mathfrak {S}} \subset {\bf
C}^n$ with ${\mathfrak {S}} =\bar\Sigma$ 
the corresponding hyperbolic equation (\ref{HYPP}) satisfies Huygens' Principle if $N$ is odd and large enough:
$N\ge 2M+3, M = \sum_{j=1}^{K} m_j$.

Conversely, if the equation (\ref{HYPP}) satisfies Huygens' Principle and all the Hadamard's coefficients are rational functions, then the potential $u(x)$ must be of the form (\ref{6.0}) for some affine locus configuration.}

{\bf Proof}. The first part of this theorem can be derived from the theorem 5.1 and the results by Yu.Berest \cite{B4} (see also \cite{BV98}). We would like, however, to present here another, more illuminating proof. It is based on a different idea which will help us to prove the second part also. 
The idea is to reduce the affine case to the linear one using the isotropic projectivisation procedure. 

The main observation is encapsulated in the following lemma.
Let $U_\nu ( x, \xi )$ \, ($\nu = 0,1,\ldots $) be some analytic functions of $2n$ variables $x=(x_1,\ldots, x_n),\,\xi=(\xi_1,\ldots, \xi_n)$ which satisfy the equations (\ref{Had}) with some potential $u(x)$. Let's define now the new functions depending on $\tilde x = (x_1,\ldots, x_n, x_{n+1}, x_{n+2})$ and 
$\tilde \xi=(\xi_1,\ldots, \xi_n, \xi_{n+1}, \xi_{n+2})$:
\begin{equation}
\label{haha}
\widetilde U_\nu ( \tilde x, \tilde \xi ) = (x_{n+1}+ ix_{n+2})^{-\nu}
(\xi_{n+1} + i\xi_{n+2})^{-\nu}  U_\nu ( \frac{x}{x_{n+1}+ ix_{n+2}}, \frac {\xi}{\xi_{n+1} + i\xi_{n+2}}) 
\end{equation}
and 
\begin{equation}
\label{hoho}
\tilde u ( \tilde x ) = (x_{n+1}+ ix_{n+2})^{-2}
u ( \frac{x}{x_{n+1}+ ix_{n+2}}). 
\end{equation}

{\bf Lemma}. {\it The relations (\ref{Had}) for $U_\nu ( x, \xi )$ and $u(x)$ are equivalent to the similar relations in $\tilde x, \tilde \xi$ for $\widetilde U_\nu ( \tilde x, \tilde \xi )$ and $\tilde u(\tilde x)$ defined by the formulas (\ref{haha}) and (\ref{hoho}).}

The proof is straightforward.

Now suppose that we have the real potential $u(x)$ related to some affine locus configuration ${\mathfrak {S}} = \bar {\mathfrak {S}} \subset {\bf C}^n$. Then the potential $ \tilde u ( \tilde x )$ defined by (\ref{hoho}) corresponds to some locus configuration
 $\widetilde {{\mathfrak {S}}} \subset {\bf C}^{n+2}$ which is exactly the result of the isotropic projectivisation defined in the previous section (see Theorem 5.4).
Thus, according to the Theorem 3.1 the corresponding Schr\"odinger operator 
$\widetilde L= -\Delta_{n+2} + \tilde u(\tilde x)$ in ${\bf C}^{n+2}$ has the BA function 
$\widetilde \psi(\tilde \xi, \tilde x)$ which is given by the formula (\ref{formula}). Therefore, 
$\widetilde \psi$ can be presented in the form
analogous to (\ref{Fi}),
\begin{equation}
\label{Fifi}
\widetilde \psi(\tilde \xi,\tilde x) = (\widetilde U_0(\tilde \xi,\tilde x)+\widetilde U_1(\tilde \xi,\tilde x)+\ldots+\widetilde U_M(\tilde \xi,\tilde x)) e^{(\tilde \xi,\tilde x)},
\end{equation}
where $\widetilde U_0=1$ and the components $\widetilde U_\nu (\tilde x,\tilde \xi)$ are homogeneous of degree $-\nu$ in $\tilde \xi$ and $\tilde x$, non-singular for $\tilde x =\tilde \xi$ and satisfy the relations  (\ref{Had}) in $\tilde x, \tilde \xi$ with the potential $\tilde u(\tilde x)$. Now let's consider their restriction for $ x_{n+1}+ ix_{n+2}=
\xi_{n+1} + i\xi_{n+2}=1$,
\begin{equation}
\label{rest}
 U_\nu ( x, \xi) =  \widetilde U_\nu (\tilde x,\tilde \xi) |_{ {x_{n+1}+ ix_{n+2}= 1} \atop
{\xi_{n+1} + i\xi_{n+2}=1}}.
\end{equation}
We claim that the formula (\ref{rest}) determines the Hadamard's coefficients for the initial potential $u(x)$. 

First of all, let's notice that this formula really determines some functions of $x, \xi$ only. This can be derived directly from the formula (\ref{formula}). Indeed, it's easy to see from the inductive procedure (\ref{phi}) that the pre-exponent in the BA function (\ref{formula}) is a linear combination of the "monomial" terms $\prod_{\alpha \in {\cal A}} (\alpha,x)^{p_{\alpha}}(\alpha,k)^{q_{\alpha}}$ with some integers  
${p_{\alpha}},{q_{\alpha}}$. Thus, $ x_{n+1}, x_{n+2},
\xi_{n+1}, \xi_{n+2}$ will enter in  $\widetilde \psi$ only as combinations  
 $ x_{n+1}+ ix_{n+2}$ and $\xi_{n+1} + i\xi_{n+2}$. This means that the coefficients $U_\nu$ defined by (\ref{rest}) indeed do not depend on 
 $ x_{n+1}, x_{n+2},\xi_{n+1}, \xi_{n+2}$. As a corollary of the homogeneity of $\widetilde U_\nu$ in $\tilde x$ and $\tilde \xi$ we may invert the formula (\ref{rest}) and obtain that $\widetilde U_\nu$ are related to $U_\nu$ by the formula (\ref{haha}). Now using the lemma we get the equations (\ref{Had}) for  
$U_\nu$.
It is clear then from (\ref{rest}) that $U_0=1$ and $U_{\nu}$ are non-singular when $x=\xi$.  
The last remark is that the procedure (\ref{rest}) gives us the real-valued functions $U_\nu$ of $x, \xi \in {\bf R}^n$ in the case when the initial potential $u(x)$ is real, ${\mathfrak {S}} = \bar {\mathfrak {S}}$.

So, for any affine locus configuration we constructed  the Hadamard's coefficients $U_\nu$ for the corresponding hyperbolic equation (\ref{HYPP}), and $U_{M+1}=0$. Applying the Hadamard's criterion, we obtain the first part of the theorem.

To prove the inverse statement, we suppose that the hyperbolic equation (\ref{HYPP}) is huygensian and has rational Hadamard's coefficients $U_\nu$ with $U_{M+1}=0$. In that case we can define the homogeneous functions $\widetilde U_\nu (\tilde x,\tilde \xi)$ by the formula (\ref{haha}). According to the lemma, they obey the equations (\ref{Had}) with the homogeneous potential (\ref{hoho}). Then in the same way as in the Theorem 6.1, we conclude that the function (\ref{Fifi}) satisfies the Schr\"odinger equation   
$\widetilde L \widetilde \psi = -\xi ^2 \psi$ with $\widetilde L= -\Delta_{n+2} + \tilde u(\tilde x)$. 
Now using the Theorem 2.2 in the same way as in the theorem 6.1 we deduce that the potential $\tilde u(\tilde x)$ must correspond to some (linear) locus configuration $\widetilde {{\mathfrak {S}}}$ of non-isotropic hyperplanes in ${\bf C}^{n+2}$. But in that case the initial potential $u(x)$ (see the formula (\ref{hoho})) will correspond to the isotropic reduction ${\mathfrak {S}}$ of $\widetilde {\mathfrak {S}}$ which should satisfy the locus equations due to the theorem 5.5. The theorem is proven.

\noindent {\it Remark}. We have assumed that the potential $u$ of the hyperbolic equation does not depend on $x_0$, 
but essentially we have used only the fact that the sequence of the Hadamard's coefficients terminates at some step $M$. 
Actually all the results of this section can be generalised formally for any equation of the form (\ref{HYP}) (even with
the complex potential), which possesses the last property. In that case the singularities of the potential should satisfy the 
locus equations in ${\bf C}^{N,1}$ with the complex Euclidean structure defined by the metrics ${\rm diag}(-1,1,\ldots ,1)$.

We conjecture that any hyperbolic equation $(\Box_{N+1} + u(x))\varphi = 0$ with terminating sequence 
of the Hadamard's coefficients has a rational potential $u(x)$ which corresponds to some locus configuration 
in ${\bf C}^{N,1}$. We have proved this under the assumption that the Hadamard's coefficients are rational.   
The proof of this conjecture would lead to the solution of the famous Hadamard's problem in the class (\ref{HYPP}).
Until now this problem is solved only when $u$ depends on one of the coordinates (K.Stellmacher, J.Lagnese \cite{SL}) 
and when $u$ is homogeneous and depends on two of the coordinates (Yu.Berest \cite{B}).

\section* {7. Some other relations and generalisations.}

{\bf 7.1.}  The Baker--Akhiezer function $\psi(k,x)$ related to an equipped configuration has
the following remarkable property: it satisfies a system of differential equations not only
in $x$ but also in $k$--variables. Corresponding bispectral property of the
one-dimensional BA function has been observed in the fundamental paper by
Duistermaat and Gr\"unbaum \cite{DG}.

Let $\psi(k,x)$ be a BA function related to some equipped configuration
${\mathfrak {A}}$, ${\mathfrak {S}}$ be the corresponding dual configuration of the poles
of the potential $u(x)$ given by (\ref{6.0}).

Let ${\cal R}$ be the ring of polynomials defined in the Theorem 5.3. Define also the dual ring ${\cal S}$ as 
the ring of all polynomials $q(x)$ in $x$,
satisfying the relations
$$
\left(\alpha_s,\frac\partial{\partial x}\right)^{2j-1}\left[q(x)\right]|_
{(\alpha_s,x)+c_s=0} \equiv 0
$$
for all $j=1,2,\ldots,m_s$ and for all the hyperplanes of the
configuration ${\mathfrak {S}}$.

{\bf Theorem 7.1.} {\it For any $p(k)\in{\cal R}$ and $q(x)\in {\cal S}$ there exist
the differential operators $L_p(x,\partial/\partial x)$ and
$M_q(k,\partial/\partial k)$ such that the BA function $\psi(k,x)$
satisfies the following bispectral problem:}
\begin{equation}
\left\{
\begin{array}{l}
L_p(x,\partial/\partial x)\psi(k,x) = p(k)\psi(k,x)\\
M_q(k,\partial/\partial k)\psi(k,x) = q(x)\psi(k,x)\\
\end{array}
\right.
\end{equation}

The existence of the operator $L_p(x,\partial/\partial x)$ is claimed in the Theorem 5.3. The existence of $M_q(k,\partial/\partial k)$ follows from the 
characterisation of $\psi$ by its analytic properties in $x$. Namely, one
can show that the BA function $\psi(k,x)$ is the unique function of the
form
$$
\psi=\frac{B(x)+\ldots}{B(x)}e^{(k,x)},
$$
where $B(x) = \prod_{s=1}^N ((\alpha_s,x)+c_s)^{m_s}$ and the dots denote
the polynomial in $x$ of a smaller degree, such that the following
conditions are fulfiled:
$$
\left(\alpha_s,\frac\partial{\partial x}\right)^{2j-1}\left[
((\alpha_s,x)+c_s)^{m_s}\psi\right]|_
{(\alpha_s,x)+c_s=0} \equiv 0
$$
for each $j=1,2,\ldots,m_s$ and $s=1,\ldots,N$. The fact that the BA
function satisfies these conditions follows from the Schr\"odinger
equation (\ref{6.5}) and the theorem 2.2.

\noindent {\bf 7.2.} Similar approach can be developed for the trigonometric versions of our Schr\"odinger operators (\ref{0.3}). As well as
in the rational case discussed in the present paper, the axiomatics of \cite{ChV1} has to be amended
in order to cover the most general case. We intend to discuss such
axiomatics in a separate paper. The
corresponding locus conditions have been described in \cite{Ch}. The bispectral property for the corresponding BA functions results in difference operators 
in the spectral parameter, which can be viewed as deformations of the
rational Ruijsenaars and Macdonald operators (see \cite{Ch1}).

\noindent {\bf 7.3.} The most of the results of this paper can be generalised to the case when
the potential $u(x)$ of the Schr\"odinger operator is a matrix-valued
function. The locus equations for that case in dimension 1 have been
described in \cite{GV}. Multidimensional case is considered in \cite{CGV}.

\noindent {\it Acknowledgements}. This work was partially supported by Russian Fundamental
Research Fund (grants 96-01-01404, 96-15-96027, 96-15-96037) and INTAS (grant 96-0770). O.Ch. was supported
also by the Royal Society postdoctoral fellowship during 1998, which is
highly acknowledged. O.Ch. and M.F. are grateful to Loughborough University,
UK, for the hospitality during the period this work was being completed.
Finally, we would like to thank Yuri Berest for extremely fruitful
discussions.

\end{document}